\documentclass[%
 reprint,
superscriptaddress,
 amsmath,amssymb,
 aps,
]{revtex4-2}

\usepackage{graphicx}
\usepackage{dcolumn}
\usepackage{bm}
\usepackage{lipsum}
\usepackage{physics}
\usepackage{multirow}
\usepackage{xcolor}
\usepackage{hyperref}
\usepackage{comment}
\usepackage{tikz}
\usetikzlibrary{quantikz2}
\usepackage[normalem]{ulem}
\newcommand{\bL}{\mathbf{L}}
\newcommand{\ba}{\mathbf{a}}
\newcommand{\bb}{\mathbf{b}}

\newcommand{\bR}{\mathbf{R}}

\newcommand{\bH}{\mathbf{H}}
\newcommand{\bS}{\mathbf{S}}
\newcommand{\bU}{\mathbf{U}}

\newcommand{\cP}{\mathcal{P}}
\newcommand{\cM}{\mathcal{M}}
\newcommand{\cG}{\mathcal{G}}
\newcommand{\bsig}{\boldsymbol{\sigma}}
\newcommand{\cD}{\mathcal{D}}
\newcommand{\brho}{\boldsymbol{\rho}}
\newcommand{\GHZ}{\ket{\text{GHZ}}}

\begin{document}

\title{Biased-noise qubits: a guide to efficient fault-tolerance using the hierarchy of errors}

\author{Diego Ruiz}
\affiliation{Alice \& Bob, 49 Bd du Général Martial Valin, 75015 Paris, France}
\affiliation{Laboratoire de Physique de l'École Normale Supérieure, École Normale Supérieure, Centre Automatique et Systèmes, Mines Paris, Université PSL, CNRS, Inria, Paris, France}
\author{J\'erémie Guillaud}%
\affiliation{Alice \& Bob, 49 Bd du Général Martial Valin, 75015 Paris, France}
\author{Christophe Vuillot}%
\affiliation{Alice \& Bob, 49 Bd du Général Martial Valin, 75015 Paris, France}
\author{Mazyar Mirrahimi}%
\email{mazyar.mirrahimi@inria.fr}%
\affiliation{Laboratoire de Physique de l'École Normale Supérieure, École Normale Supérieure, Centre Automatique et Systèmes, Mines Paris, Université PSL, CNRS, Inria, Paris, France}

\date{\today}

\begin{abstract}
Qubits with strongly biased noise, in which phase-flip errors are orders of magnitude more frequent than bit-flips, arise both naturally, as in electron and nuclear spins, and by engineering, as in stabilized cat qubits. This noise structure holds the promise of reducing the daunting hardware overhead of fault-tolerant quantum computing, but exploiting it requires physical operations that do not convert frequent phase-flips into rare bit-flips. In this review, we analyze the most prominent fault-tolerant protocols for biased-noise qubits, organized according to the available set of such bias-preserving operations. When this set is restricted to the C\(Z\) gate together with preparation and measurement in the \(X\) basis, we show that the complexity of the required syndrome extraction gadgets essentially cancels the benefit of the noise bias: at experimentally relevant error rates, one may as well ignore the bias and rely on standard error correction designed for depolarizing noise. The situation changes drastically when a bias-preserving C\(X\) gate is available: the hierarchy of errors can then be reflected in the structure of the code, with frequent phase-flips corrected by a dedicated high-threshold code and rare bit-flips by concatenation with a high-rate code. The same hierarchy also enables hardware-efficient preparation of magic states. Finally, as a bias-preserving C\(X\) is forbidden in naturally biased platforms and challenging in engineered ones, we present a measurement-based architecture in which a high-fidelity quantum non-demolition readout of multi-qubit Pauli \(Z\) operators takes its place, extending these overhead reductions to a much broader range of physical platforms.
\end{abstract}

\maketitle

\tableofcontents

\section{\label{sec:intro}Introduction}



 

Quantum error correction (QEC) is widely viewed as essential to reach a regime in which quantum computers can provide a useful advantage over their  classical counterparts~\cite{eisert2025mind}. In the seminal work by Peter Shor~\cite{shor1995scheme}, it was argued that, to correct the effect of decoherence on single qubits, it suffices to find quantum error-correcting codes that correct bit-flip (Pauli $X$) and phase-flip (Pauli $Z$) errors. This important observation rapidly led to many constructions of QEC codes~\cite{shor1995scheme,calderbank1996good,steane1996error,Laflamme-5qubit-96,Gottesman-PRA-96,kitaev1997quantum}. Among these constructions, the 2D toric code~\cite{kitaev1997quantum} or its planar version, the surface code~\cite{bravyi-kitaev-surface-1998},  became increasingly popular over the last decades. By encoding the logical information of a quantum bit in the many-body state of a $d$ by $d$ lattice of physical qubits, and by measuring weight-4 Pauli $X$ and Pauli $Z$ parity checks, this code corrects up to $\lfloor(d-1)/2\rfloor$ bit-flip and phase-flip errors. This code is resistant to an interestingly high physical error rate. More precisely, whenever the probability of errors per physical operation is lower than a threshold of about $1\%$~\cite{fowler2012surface}, this code suppresses the logical error exponentially when increasing the code distance $d$ (the minimal number of undetected errors)~\cite{kitaev2003fault}. Furthermore, the 2D layout of the code with local stabilizers makes this theoretical construction relevant for actual implementations within the framework of solid-state quantum computing platforms such as superconducting qubits. These interesting features of the surface code have led to a thorough investigation of its experimental implementation~\cite{Krinner-Nature-2022,Zhao-surface-22,google2023suppressing}. In particular, the exponential suppression of the errors below the error correction threshold was recently demonstrated in~\cite{google2025quantum}.

Despite this tremendous progress in the realization of QEC codes, the overhead of quantum error correction remains daunting and continues to be a major roadblock for quantum computing~\cite{gidney2025factor}. It is estimated that reaching algorithmically relevant error rates would require an overhead in the order of hundreds to thousands of physical qubits per logical qubit when relying on physical error rates within the reach of the  current experiments~\cite{beverland2022assessing}. Note furthermore that this overhead only concerns the preservation of quantum information in a quantum memory and the fault-tolerant operations come with an extra overhead. Although recent theoretical developments point towards significant overhead savings~\cite{litinski2019magic,gidney2024magic}, the high-fidelity preparation of magic states for fault-tolerant non-Clifford gates remains particularly hardware-consuming.

The experimental progress and the advent of new qubit designs could, in principle, reduce this overhead by reaching physical error rates that are significantly below the threshold of the surface code.  In parallel to such experimental efforts, a
number of theoretical proposals have investigated possible shortcuts to QEC and fault tolerance. 

One such shortcut is through designing better QEC codes, and in particular the so-called quantum Low-Density Parity-Check (qLDPC) codes~\cite{breuckmann2021quantum}. These codes encode many logical qubits in many physical qubits and  by sharing physical resources among encoded logical qubits lead to an enhanced encoding rate. The intensive research efforts in this direction recently led to the advent of ``good'' qLDPC codes~\cite{panteleev2022asymptotically,leverrier2022quantum,dinur2023good} with a non-vanishing encoding rate and a distance linearly increasing with the code size. In parallel to these theoretical results, small instances of qLDPC codes with performances better than the surface code were also discovered~\cite{bravyi2024high,Shaw-PRL-25,Ye-quantum-25,Yang-npjQI-26}, revealing the practical interest of this route. However, ensuring a better performance than the surface code with these codes necessitates to go beyond the 2D local connectivity~\cite{bravyi2010tradeoffs}. Within the context of superconducting qubits, this implies new technological developments such as high-quality long-range couplers. Furthermore, even in reconfigurable atomic platforms such as  atom arrays in optical tweezers~\cite{bluvstein2024logical} and trapped ions~\cite{Helios-Quantinuum-26}, this enhanced connectivity requires longer coherent atomic transports, effectively increasing the physical error rates and also the error correction cycle time. 

A second type of shortcut corresponds to hardware-level error correction. The realm of bosonic error correction falls within this route~\cite{joshi2021quantum}. The large Hilbert space of a single quantum harmonic oscillator, or a few such systems, provides the redundancy needed for error correction, thus replacing a many-qubit system. By encoding the quantum information on the so-called grid states of light, the  GKP (Gottesman-Kitaev-Preskill) code provides protection against small shifts in the phase space. This is the class of translation-symmetric codes. In contrast, the cat codes~\cite{cochrane1999macroscopically,leghtas2013hardware} and binomial codes~\cite{michael2016new} are subclasses of the so-called rotation symmetric codes~\cite{Grimsmo-PRX-20}. These codes can provide an efficient protection against dominant physical noise channels such as photon loss or dephasing. To reach algorithmically relevant logical error rates, they however need to be concatenated with discrete-variable codes such as the surface code~\cite{fukui2017analog,vuillot2019quantum,guillaud2019repetition,Darmawan-PRXQ-21,chamberland2022building}. The hardware-level bosonic error correction  ensures a significant overhead reduction for this concatenation. Such ideas have recently been also extended to other high-dimensional quantum systems such as large nuclear spins~\cite{gross2021,gross2024,omanakuttan2024fault,Kruckenhauser-2025}.

The simplest instance of such bosonic codes, the two-component cat code, can be autonomously stabilized by  two-photon driven dissipation~\cite{Wolinsky-Carmichael-88,mirrahimi2014dynamically}. Stabilized cat qubits can be seen as biased-noise qubits, where bit-flip errors are exponentially suppressed with the average number of photons~\cite{lescanne2020exponential,reglade2024quantum,rousseau2025enhancing}. While other types of qubits such as spins, or the hyperfine clock transition qubits in atomic systems, benefit also from noise bias, there are some major differences between them, in particular, with respect to the physical operations that preserve the noise bias. In parallel to the developments of quantum error-correcting codes and concepts of fault tolerance, a number of theoretical proposals have discussed ways to tailor these protocols for biased-noise qubits. The advent of biased-noise cat qubits with new capabilities intensified this research and led to many new ideas.

This review examines the most prominent fault-tolerant protocols for quantum memory and quantum computation in the context of biased noise. More precisely, we consider qubits where the probability of phase-flip (Pauli $Z$) errors, $p_Z$, is orders of magnitude higher than the probability of bit-flip type (Pauli $X$ and $Y$) errors, $p_X$ and $p_Y$. More precisely, we call $\eta=p_Z/(p_X+p_Y)\gg 1$ the bias value and, we assume for the sake of simplicity that $p_X=p_Y$.  We start the review, in Section~\ref{sec:bias}, by an introduction to such biased-noise qubits, both of natural type and of engineered type. We also discuss the concept of bias-preserving operations as building blocks of noise-tailored fault-tolerant protocols. After introducing a set of admissible bias-preserving gates, we argue that not all such gates are implementable with all biased-noise qubits. 

Based on the elementary set of bias-preserving physical operations that are implemented within a biased-noise platform, we discuss optimized architectures for protecting quantum information in a quantum memory and also the implementation of fault-tolerant operations. We divide such architectures in three categories presented in Sections~\ref{sec:CZ},~\ref{sec:CNOT}, and~\ref{sec:ZZZ meas}. 

In Section~\ref{sec:CZ}, we consider the case of a minimal elementary set only consisting of $\{\cP_{\ket{+}},\cM_X,\text{C}Z\}$, where $\cP_{\ket{+}}$ stands for the preparation of $\ket{+}=(\ket{0}+\ket{1})/\sqrt{2}$ and $\cM_X$ stands for the measurement of the Pauli $X$ operator. Finally, C\(Z\) stands for a two-qubit controlled-\(Z\) operation. The analysis of Section~\ref{sec:CZ} points towards the very small (or even no) benefit of the natural or engineered suppression of bit-flips for quantum error correction purposes. In other words, if the elementary set of bias-preserving operations is limited to the above set, we cannot really exploit the  absence or rarity of bit-flips: we may forget about the noise bias and treat such qubits as  those exposed to a more balanced noise of the same order. 

We then argue in Section~\ref{sec:CNOT} that the situation changes drastically when we add a bias-preserving C\(X\) (also known as CNOT) operation to this elementary gate set. We provide a thorough comparison between the most prominent error correction strategies based on the elementary bias-preserving gate set  $\{\cP_{\ket{0}},\cP_{\ket{+}},\cM_Z,\cM_X,\text{C}Z,\text{C}X\}$, and come to the conclusion that if the noise bias is significant (e.g. dissipative cat qubits), it is most convenient to design a code for correcting only the phase-flips and then concatenate with a bit-flip code.  We also discuss noise-tailored fault-tolerant operation designs that are based on this elementary gate set and in addition either a  bias-preserving $ \text{CC}X$ (Toffoli) operation or  bias-non-preserving operations $\{X^{\pm 1/4}\}$. 

Noting that such a bias-preserving C\(X\) operation cannot be implemented in many biased-noise platforms, we focus in Section~\ref{sec:ZZZ meas} on a third type of architecture with a primitive consisting of high-fidelity Quantum Non-Demolition (QND) readout of the multi-qubit Pauli $Z$ operator. More precisely, we demonstrate that all the constructions of Section~\ref{sec:CNOT} can be extended to the case where the set of elementary operations is given by bias-preserving operations $\{\cP_{\ket{+}},\cM_X,\text{QND-}\cM_Z,\text{QND-}\cM_{Z^{\otimes 3}},\text{C}Z\}$ together with bias-non-preserving operations $\{X^{\pm 1/4}\}$.

\section{Biased-noise qubits and bias-preserving operations}\label{sec:bias}

The relaxation time $T_1$ and the coherence time $T_2$ serve as reliable proxies for assessing a qubit's performance. Since relaxation to the $\ket{0}$ state necessarily destroys the qubit's phase coherence, this imposes $T_2 \le 2T_1$. However, in most qubit implementations, noise sources are not symmetric. The basis states of a qubit are typically encoded in the eigenstates of a Hamiltonian, which makes the relaxation time $T_1$ generally longer than the coherence time $T_2$. While relaxation arises from energy exchange between the qubit and its environment, dephasing results from entanglement with the environment. This can lead to a coherence time that is not $T_1$-limited and can be significantly shorter than the relaxation time~\cite{feder2025fluorescent,steinacker2025industry, huang2024high}. This is also commonly observed for neutral atoms~\cite{evered2023high} and trapped ions~\cite{ruster2016long,goodwin2016resolved, tan2015multi}.

Throughout this section, we consider two types of noise-biased qubits: 1- those with a natural noise bias due to  rarer relaxation events, 2- those with engineered noise bias where the bit-flip errors are actively suppressed through a hardware-level error correction.  For both these types of qubits, we will review the set of quantum operations that can be implemented while preserving the noise bias. These operations include state preparations, measurements, and unitary gates. We will see in the next sections that the significance of the overhead reduction for fault-tolerance directly depends on this  set of elementary bias-preserving operations. 

\subsection{Natural noise-bias}\label{ssec:natural}

Consider for instance an electron spin in a magnetic field $B_0\vec{e}_z$. The dynamics of the spin is determined by the Zeeman Hamiltonian  
$$
\bH=\gamma_e B_0\bS_z+\gamma_e\vec{\mathbf{B}}_{\text{noise}}(t)\cdot \vec{\bS},
$$
where $\vec{\bS}=(\bS_x,\bS_y,\bS_z)$ represent the spin operators $\bS_{x,y,z}=\hbar\bsig_{x,y,z}/2$, $\gamma_e$ is the electron's gyromagnetic ratio, and  $\vec{\mathbf{B}}_{\text{noise}}$ represents the magnetic noise which can be classical or quantum. This naturally defines a spin qubit where the $T_1$ relaxation requires a transverse magnetic noise (perpendicular to the $Z$ axis) close to resonance with the Larmor frequency given by $\gamma_eB_0$. While such a relaxation mechanism can perhaps be avoided by careful engineering of the environment of the electron spin or by an appropriate choice of the Zeeman  field, parallel magnetic noise at any frequency can lead to a direct unavoidable dephasing of the qubit.  Moreover, note that the transverse fields could also lead to such a dephasing through a second order effect, a random stark shift in qubit's frequency. This natural noise asymmetry is further enhanced when we switch to quantum systems with a weaker interaction with their environment, as the filtering of the resonant relaxation process becomes easier. This is, for instance, the case of nuclear spins that admit a smaller gyromagnetic ratio by a factor of about a thousand. On the other side of the spectrum, for qubits that are more strongly coupled to the environment, such as charge qubits in semiconductors or superconductors, the asymmetry is washed out by the difficulty in suppressing the relaxation processes. 

Note that, even if a qubit exhibits a significant noise bias at rest, bias-preserving operations are required to fully leverage this property in a quantum processor~\cite{aliferis2008fault}. A bias-preserving operation is defined as an operation that maps phase-flip errors exclusively to phase-flip components. A necessary condition for an $n$-qubit unitary gate $\bU$ to be  bias-preserving is that  $\bU \mathcal{Z}_n \bU^\dagger \subseteq \mathcal Z_n $ where $\mathcal Z_n$ is the subgroup of the Pauli group defined as $\mathcal Z_n = \{\omega P_1 \otimes \dots \otimes P_n\ |\ \omega \in \{\pm 1, \pm i \}  , P_j \in \{I,Z\} \}$~\cite{guillaud2021repetition}. This reduces the group of admissible single-qubit unitary gates to all rotations around the $Z$-axis $\bR_{Z}(\theta)$, as well as the single-qubit Pauli gates $X$ and $Y$. Among the multi-qubit gates, the operations
C$Z$, C$X$, CC$X$ and CC$Z$ are all admissible. But, clearly the Hadamard gate and $X$ or $Y$ rotations with an angle that is not a multiple of $\pi$ are excluded.

\begin{figure}
    \centering
    \includegraphics[width=\columnwidth]{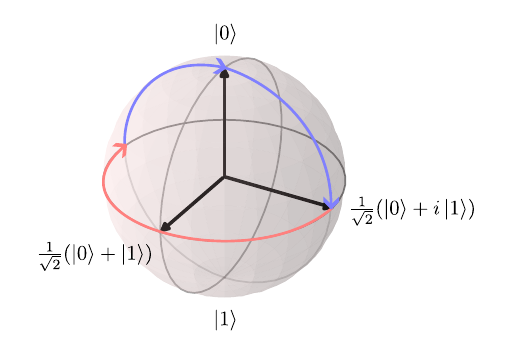}
    \caption{Implementation of an $X$ gate. Starting from the state $\ket{0}$, a continuous rotation is applied targeting the state $\ket{1}$ (blue arrow). However, when the qubit is in a superposition of the states $\ket{0}$ and $\ket{1}$, it is sensitive to phase-flip errors (red arrow). Such phase-flip errors prevent the qubit from reaching the target state $\ket{1}$, thereby converting phase-flip errors into bit-flip errors.}
    \label{fig:X_nonbiaspreserving}
\end{figure}

Although a gate may be admissible following the above rule, it might lack a bias-preserving implementation. For example, if a Pauli $X$ gate is implemented by continuously rotating around the $X$-axis of the Bloch sphere from an angle 0 to $\pi$, a phase-flip error occurring during the rotation can be mapped to a bit-flip error, as illustrated in Figure~\ref{fig:X_nonbiaspreserving}. Indeed, the set of bias-preserving operations depends on the qubit platform and the way these gates are implemented. One crucial operation, that is a central ingredient of fault-tolerant architectures, is the C\(X\) or CNOT gate. It is in particular used to perform the parity checks of the error correcting codes, by mapping the parity information of data qubits on the measurement ones. Note that while in a number of platforms the C\(Z\) gates are more natural than C\(X\), they need to be combined with single-qubit Hadamard gates, or $\pi/2$ rotations around $X$ or $Y$ axis, to perform the required parity checks. But such single-qubit gates are not admissible as bias-preserving operations. Finally, note that while C\(X\) is admissible, similarly to the Pauli $X$ gate, it lacks a bias-preserving implementation when restricted to an evolution in  strictly  two-dimensional Hilbert spaces of qubits~\cite{guillaud2019repetition}. 

This argument excludes the C\(X\) gate as a relevant gate for most biased-noise platforms as when used in an error correction circuit, it depolarizes the noise and therefore washes out the natural suppression of bit-flips~\cite{gravier2025}. In contrast, the gates that are diagonal in the $Z$-basis are generally much more likely to allow for a bias-preserving implementations. For instance, the references~\cite{Brito_2008,Aliferis_2009} proposed a bias-preserving implementation of the C\(Z\) gate in a biased-noise superconducting flux qubit. Such constructions can be more or less easily generalized to other types of biased-noise qubits. 

In Section~\ref{sec:CZ}, we review the fault-tolerant constructions for a biased-noise qubit with a small set of elementary bias-preserving operations given by $\{\text{C}Z,\cP_{\ket{+}},\cM_X\}$. We will see that the restriction of physical-level operations to this set strongly limits the interest of biased noise. More precisely, with this restricted set, the hardware overhead of error correction with biased-noise qubits is not significantly better than in the case of  qubits with the same level of depolarizing noise. In the absence of a bias-preserving C\(X\), we will see in Section~\ref{sec:ZZZ meas} that an alternative hardware-efficient approach is possible using high-fidelity quantum non-demolition measurements of multi-qubit Pauli $Z$ measurements. As opposed to the bias-preserving C\(X\), such measurements are not forbidden for naturally noise-biased qubits. In the next  paragraph, following our recent work~\cite{vuillot2026}, we briefly discuss possible physical implementations of such an operation with nuclear spin qubits. Similar protocols could be designed for other naturally biased-noise qubits. 

The construction of~\cite{vuillot2026} is based on the hyperfine coupling of nuclear spins to a single electron spin. By driving the electronic spin with an appropriate multi-tone drive, it is possible to map the three- or four qubit $Z$ parity of nuclear spins to the electron spin state. The electron spin then needs to be measured without revealing information on individual nuclear spins. Indeed, revealing such information leads to a dephasing of individual nuclear spins. Some readout protocols, such as optical readout of NV centers coupled to $^{13}C$ nuclear spins~\cite{Jiang-PRL-2008,Pfaff-NatPhys-2013}, or electrical readout of  quantum dots coupled to nuclear spins~\cite{Hensen_Nat_Nano,steinacker2025industry}, could allow for such readout, preserving the coherence of nuclear spins. As it will be seen in the constructions of Section~\ref{sec:ZZZ meas}, the proxy for the bit-flip error probability in this QND readout protocol is the infidelity of the measurement. In a truly QND readout, this infidelity can be arbitrarily reduced by repeating the measurements. Therefore, the bit-flip error probability is ultimately limited by the deviations from the QND nature of these readout protocols. We refer to the discussion in~\cite{vuillot2026} and~\cite{Travesedo-2025} for approaches to push this limit.

\subsection{Engineered noise-bias}\label{ssec:engineered}

In this subsection, we discuss another type of noise-bias that can be seen as an engineered and autonomous error correction at the hardware level. The principle behind such kind of protection is that, by benefiting from the large Hilbert space of a bosonic mode~\cite{cochrane1999macroscopically,leghtas2013hardware} or a large nuclear spin~\cite{omanakuttan2024fault}, we can encode the quantum information in states that are well-separated in the phase space. Furthermore, it is possible to engineer a certain dissipative  mechanism that autonomously stabilizes the two-dimensional manifold of the encoding states and as such removing the possibility of jumps between the computational states, also known as bit-flips~\cite{mirrahimi2014dynamically,Kruckenhauser-2025,Rojkov-PRX-2026}. These so-called bosonic or spin cat qubits can further be operated while preserving the suppression of bit flips. More interestingly, their large Hilbert space can be exploited to perform bias-preserving operations that are not allowed with naturally biased-noise qubits. Throughout this subsection, we will introduce the particular case of bosonic cat qubits stabilized by two-photon dissipation and we refer to the above references for some other types of cat qubits.

Before presenting the dissipative bosonic cat qubits, we would like to emphasize the existence of another type of cat qubit confinement, called the Kerr or simply the Hamiltonian confinement~\cite{puri2017engineering}. Through two-photon driving of a Kerr resonator, a  Hamiltonian with a two-fold degenerate eigenspace that is separated by a gap from the rest of the spectrum is engineered. While this gap protects the qubit from leakage induced by slow noise (interaction Hamiltonians that are weaker than the gap), the thermal excitation and Markovian dephasing can still lead to a leakage out of the code space and as a second order effect  bit-flip type errors in the cat qubit manifold~\cite{gautier2022combined,Su-PRA-2025}. As such, to avoid the leakage and to retrieve the  exponential suppression of bit-flips it needs to be combined with an appropriate dissipative mechanism~\cite{putterman2022stabilizing,gautier2022combined}. In absence of such a dissipative mechanism, the Kerr cat qubit can be used as qubits with a moderate noise bias. 

The key principle of bosonic qubits is that quantum information is stored in a delocalized manner across the bosonic mode's phase space, whereas the dominant errors in a harmonic oscillator act locally in phase space. In particular, when the dissipators or Hamiltonians are expressed as polynomials of the ladder operators, the Wigner function of the density matrix evolves through continuous local deformations. Cat qubits encode information in spatially separated coherent states of a harmonic oscillator, with this separation providing protection against transitions between the two. 

\begin{figure}
    \centering
    \includegraphics[width=\columnwidth]{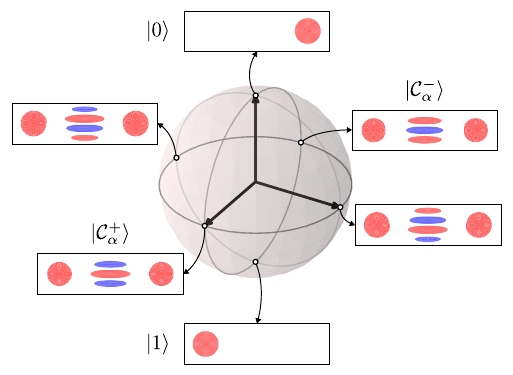}
    \caption{Bloch sphere of a cat qubit.}
    \label{fig:cat_bloch}
\end{figure}

More precisely, the information is encoded in the coherent states $\ket{\alpha}$ and $\ket{-\alpha}$ of a harmonic oscillator, where $\ket{\alpha} = e^{-|\alpha|^2/2} \sum_{n \ge 0} \frac{\alpha^n}{\sqrt{n!}} \ket{n}$. The qubit's states are defined as 
\begin{equation}
\ket{\pm} = \ket{\mathcal C ^\pm_\alpha} =  \frac{\ket{\alpha} \pm \ket{-\alpha}}{\sqrt{2(1 \pm e^{-2|\alpha|^2})}}.
\end{equation}
The computational basis states are then expressed as $\ket{0} = \ket{\alpha} + \mathcal O (e^{- 2 |\alpha|^2})$ and $\ket{1} = \ket{-\alpha} + \mathcal O (e^{- 2 |\alpha|^2})$, as illustrated in Figure~\ref{fig:cat_bloch}. As, for a large number of photons, the states $\ket{0}$ and $\ket{1}$ are well separated in phase space, cat qubits can be made robust to bit-flip errors when subject only to local errors such as photon loss, dephasing, thermal noise, etc. This potential for bit-flip suppression can be exploited using a nonlinear dissipative mechanism called the two-photon dissipation~\cite{Wolinsky-Carmichael-88,mirrahimi2014dynamically}. This corresponds to the engineering of a two-photon exchange Hamiltonian between a high-Q storage mode $\ba$ and a driven lossy buffer mode $\bb$. In an appropriate rotating frame, this corresponds to an interaction Hamiltonian of the form 
$$
\bH_{2\text{ph}}=g_{2\text{ph}}\ba^{2}\bb^\dag+g_{2\text{ph}}^*\ba^{\dag 2}\bb+\epsilon_d \bb^\dag+\epsilon_d\bb.
$$
By choosing the amplitude of the drive $\epsilon_d=-g_{2\text{ph}}\alpha^2$, this Hamiltonian can also be written as $\bH_{2\text{ph}}=g_{2\text{ph}}(\ba^2-\alpha^2)\bb^\dag+$h.c.. Assuming the loss rate $\kappa_b$ of the buffer mode to exceed this two-photon interaction strength, we can adiabatically eliminate the buffer mode and write an effective master equation for the storage mode~\cite{mirrahimi2014dynamically}:
$$
\frac{d}{dt}\brho=\kappa_{2\text{ph}}\cD[\ba^2-\alpha^2]\brho,
$$
where $\cD[\bL]\brho=\bL\brho\bL^\dag-\bL^\dag\bL\brho/2-\brho\bL^\dag\bL$, and $\kappa_{2\text{ph}}\approx 4|g_{2\text{ph}}|^2/\kappa_b$. The two coherent states $\ket{\pm\alpha}$ being in the kernel of the Lindblad operator, this master equation admits as steady state the 2D manifold spanned by these coherent states. While this engineered dissipation stabilizes the cat qubit manifold and removes any leakage out of this manifold, the perturbations to these dynamics, such as undesired (single photon) loss of the storage mode, its thermal excitation and dephasing, could lead to effective errors in the qubit subspace. However, these errors are mainly of the phase-flip form. Indeed, as demonstrated in~\cite{lescanne2020exponential,reglade2024quantum}, the bit-flip  errors are exponentially suppressed with the average number of photons in the cat state $|\alpha|^2$. Note that this amplitude is a simple knob in the experiments and can be tuned via the amplitude of the resonant buffer drive $|\epsilon_d|$. However, this increase in the average photon number in the cat state is not for free and leads to a linear increase in the phase-flip error rate.  This favorable scaling (exponential suppression of bit-flips and linear increase of phase-flips) leads to very large noise biases: 7 or 8 orders of magnitude in some recent experiments~\cite{reglade2024quantum,rousseau2025enhancing}.

Interestingly, beyond such large noise biases, the cat qubits are also compatible with a bias-preserving implementation of all the gates of the admissible set as defined in the previous subsection. First, the bias-preserving implementation of the gates that are diagonal in the computational basis, such as single-qubit rotations around the $Z$ axis or the two-qubit C\(Z\) gate, can be ensured via Zeno dynamics~\cite{mirrahimi2014dynamically}. By weakly and resonantly driving the storage mode while it is stabilized in the cat manifold under the two-photon driven dissipation, one can implement in a bias-preserving  manner any single-qubit rotation around the $Z$ axis~\cite{touzard2018coherent,reglade2024quantum,rousseau2025enhancing}. 

Remarkably, it is also possible to implement bias-preserving \(X\), C\(X\) and CC\(X\) operations on confined cat qubits as discussed in~\cite{Puri-PRX-2019,guillaud2019repetition,puri2020bias}. Indeed, the qubit being defined as a two-dimensional subspace of an infinite dimensional Hilbert space, it is possible to bypass the no go argument of the previous subsection. Indeed, the Pauli \(X\) gate can be implemented by rotating the encoding coherent states of the cat qubit in the phase space of the harmonic oscillator. This can be done, for instance, by modifying the two-photon driven  dissipation Lindblad operator to $\ba^2-\beta(t)^2$, where $\beta(0)=\alpha$ and $\beta(T)=-\alpha$~\cite{guillaud2019repetition}. By assuming an adiabatic evolution, the cat qubit follows an encoding in the manifold $\text{span}\{\ket{\pm\beta(t)}\}$. As far as  the states $\ket{\pm\beta(t)}$ remain separated in the phase space during the evolution, the bit-flip probability remains exponentially small. Another approach is to turn off the two-photon dissipation and shift the cat mode frequency by $\Delta$. After a time duration of $\pi/\Delta$, the cat qubit has rotated by $\pi$ in the phase space and we can turn on the two photon dissipative confinement again. The dephasing accumulated during this free evolution, if it is not too strong, will get corrected with this stabilization~\cite{Puri-PRX-2019}. These ideas can be extended to C\(X\) and CC\(X\) gates but come with some complications. First, they require the engineering of new interaction Hamiltonians. Second, the gate needs to be slow as non-adiabatic effects could lead to  phase-flip (and even some bit-flip) errors in cat qubits. Finally, the undesired nonlinearities such as Kerr effects and other  dephasing sources could lead to a distortion of the cat states during the free evolution which leads to an enhanced bit-flip rate after stabilization. Some recent theoretical results deal with solutions to such limitations~\cite{xu2022engineering,gautier2022combined,gautier2023designing}.

Similarly to naturally noise-biased qubits, one alternative to such bias-preserving C\(X\) gate is the high-fidelity QND readout of multi-qubit Pauli \(Z\) operators.  In our recent publication~\cite{vuillot2026}, we propose a protocol to realize such measurements and we discuss how this solves some of the major limitations with the implementation of a bias-preserving C\(X\).

\section{\label{sec:CZ}C\(Z\)-based architecture}

While a number of  publications~\cite{gourlay2000concatenated, ioffe2007asymmetric, sarvepalli2008asymmetric} had considered the realization of codes with asymmetric distances in $X$ and $Z$ to benefit from the natural noise bias existing in many qubits, the reference~\cite{aliferis2008fault} was the first one that discussed thoroughly the question of bias-preserving operations. Indeed, the authors rightfully argued that the operations required, for instance, in an error correction protocol can depolarize the biased noise of the qubit. Thus, such asymmetric codes are not necessarily helpful in a straightforward manner, as the phase-flip errors can be converted to bit-flip ones during the parity-check measurements for error correction.

This article~\cite{aliferis2008fault} and the follow-up work~\cite{brooks2013fault} then considered a minimal set $\mathcal G = \{\text{C}Z, \mathcal P_{\ket{+}}, \mathcal M_X\}$ of elementary operations that are easily compatible with the noise bias. The two-qubit C\(Z\) gate is diagonal in the computational basis and as such it commutes with phase-flip errors. Note, however, that this is  not enough for the gate to be bias-preserving as one should also ensure that the phase-flips during the implementation of the gate cannot be converted to bit-flips. The reference~\cite{Aliferis_2009} discusses one such implementation in the case of biased-noise superconducting qubits. The preparation $\mathcal{P}_{\ket{+}}$ and the measurement $\mathcal{M}_X$ are naturally compatible with the noise bias. Indeed, as the state $\ket{+}$ is invariant under bit-flip errors, any error in its preparation propagates in the logical circuits as phase-flip errors. Similarly any assignment error in the measurement $\mathcal{M}_X$ can be seen as a perfect measurement preceded by a phase-flip error. 

Based on this minimal gate set, the reference~\cite{aliferis2008fault} builds a set of fault-tolerant gadgets on a repetition code ensuring a first level of phase-flip error correction. This approach is then extended in~\cite{brooks2013fault} to an asymmetric Bacon-Shor code for correcting both phase-flips and bit-flips. In the following subsections, we briefly overview the main gadgets required for a quantum memory and a universal quantum processor, and we discuss the performance of this approach through numerical simulations.

\subsection{Biased-noise error correction}\label{ssec:knill}

\begin{figure}[h!]
    \centering
    \includegraphics[width=\columnwidth]{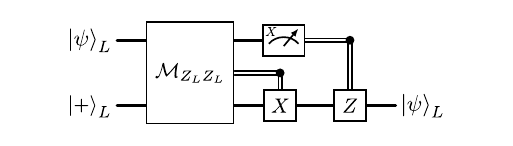}
    \caption{Knill error correction using a joint $ZZ$ logical measurement. Each line here represents a logical qubit encoded in a code block. The circuit teleports the encoded logical qubit $\ket{\psi}_L$ from the first logical block to the second. The fault tolerance of various operations in this protocol, as discussed in the text,  ensures the error correction property of this teleportation protocol. Throughout this review, the thick lines  in  logical circuits (like in this plot) represent logical qubits whereas thin lines (like in Fig.~\ref{fig:Aliferis-circuit}) represent physical qubits.}
    \label{fig:Aliferis Knill}
\end{figure}

Two approaches can be pursued for error correction with the set of bias-preserving gate $\mathcal G$. The first approach is the one introduced in~\cite{aliferis2008fault,brooks2013fault} relying on a Knill-style error correction. The second approach is discussed in our recent publication~\cite{vuillot2026} and uses a Shor-style error correction, where $XX$ stabilizer measurements are compiled  using C\(Z\) operations.   

\textit{Knill-style approach.}---The quantum error correction relies on the use of one-bit teleportation protocol depicted in Fig.~\ref{fig:Aliferis Knill}. The implementation of this teleportation protocol, at the logical level of a repetition code, provides a Knill-style phase-flip error correction gadget~\cite{Knill-2005}. Indeed, in the case of the repetition code, we note that the final logical $X$ measurement of the first block is done transversally by applying individual $\mathcal{M}_X$ measurements on each encoding qubit and by performing a majority vote. This majority vote addresses the accumulated phase-flip errors in $\ket{\psi_L}$ that propagate through the joint logical $Z_LZ_L$ measurement  and alter these individual $\mathcal{M}_X$ outcomes. In the case of the asymmetric Bacon-Shor code, the $X$ parity of each column is first calculated and then the logical $X$ value is given by a majority   vote over the results of these column parities. Furthermore, the fault tolerance of the logical non-destructive $\mathcal{M}_{Z_LZ_L}$ ensures the protection against (rarer) bit-flip errors as will be discussed later.

\begin{figure}
 \centering
    \includegraphics[width=\columnwidth]{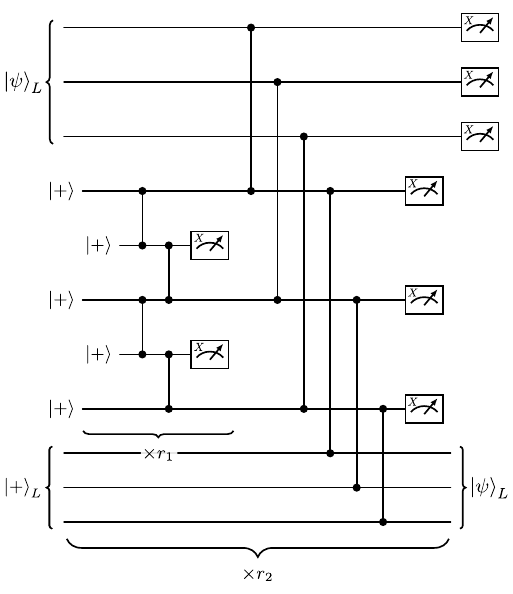}
    \caption{Knill-style error correction circuit with bias-preserving C\(Z\) gates as presented in~\cite{aliferis2008fault}. Measurement of a high-weight $Z$ operator associated with the $Z_LZ_L$ parity required for the 1-bit teleportation protocol of Fig.~\ref{fig:Aliferis Knill} is ensured using bias-preserving $\text C Z$ gates. Here each line represent a physical qubit. The first step is to prepare a $\GHZ$ state. This is performed by preparing $n$ data qubits in $\ket{+}^{\otimes n}$ and measuring neighbouring $ZZ$ stabilizers of the state using $n-1$ ancilla qubits (here $n=3$). This measurement is noisy and is thus repeated $r_1$ times. This $\GHZ$ state is then used to measure the weight-$2n$ $Z$ operator (effectively measuring $Z_LZ_L$ on two code blocks) with the application of transversal $\text C Z$ gates. These measurements being noisy, we need to repeat the whole process $r_2$ times. }
    \label{fig:Aliferis-circuit}
\end{figure}

The central ingredient in this teleportation protocol, and all other gadgets proposed for universal quantum computation, is a fault-tolerant QND measurement of high-weight $Z$ operators, as presented in Figure~\ref{fig:Aliferis-circuit}~\cite{aliferis2008fault,brooks2013fault}. First, a $\GHZ = (\ket{0}^{\otimes n} + \ket{1}^{\otimes n})/\sqrt{2}$ state is prepared starting from $\ket{+}^{\otimes n}$ and measuring the state stabilizers $ZZ$ for $r_1$ rounds. The measurement outcomes are used in a minimum-weight perfect matching decoder to identify the entangled state that has been prepared. This entangled state can be modified to a $\GHZ$ by applying physical $X$ gates on appropriate qubits, but we can also skip the physical realization of these gates by simply keeping track of them in the Pauli frame. Then C\(Z\) gates are performed transversally between the $\GHZ$ state and the qubits in the support of the $Z$ operator to be measured. The result of the high-weight $Z$ measurement is then computed by taking the product of $X$ measurements on all qubits of the $\GHZ$ state. In Fig.~\ref{fig:Aliferis-circuit}, the $\GHZ$ state is consumed to measure a weight-$2n$ $Z$ operator associated to the joint $Z_LZ_L$ measurement between two repetition codes as required by the Knill error correction shown in Fig.~\ref{fig:Aliferis Knill}. Note that any phase-flip error $Z$ on the qubits of the $\GHZ$ state flips the measurement outcome. To obtain a reliable measurement, the whole process is repeated $r_2$ times to reach the desired fidelity. Note furthermore that any bit-flip error on the data qubits also leads to a measurement error. In the case of the repetition code this probability of a single bit-flip gives a lower bound for the failure probability of  high-weight \(Z\) readout. However, in the case of the asymmetric Bacon-Shor code~\cite{brooks2013fault}, the measurement is also repeated on all rows of the code and a majority vote is considered. Therefore, for the failure of the high-weight \(Z\) readout, one needs a bit-flip event in a majority of rows.

In order to finish the presentation of this protocol, we note that besides the destructive measurement of the logical $X$ operator, and the non-destructive measurement of $Z_LZ_L$, we also need a fault-tolerant preparation of the logical  $\ket{+}_L$ state. This is straightforward in the case of the repetition code, and can be done in a transversal manner by preparing  individual $\ket{+}$ states in each data qubit. For the asymmetric Bacon-Shor code, this boils down to preparing short $\GHZ$ states for each column of the code following the same procedure as the ancilla $\GHZ$ state for high-weight \(Z\) readout. 

Through  the numerical simulations of subsection~\ref{ssec:CZnumerics}, we will see that all this overhead required for these fault-tolerant gadgets strongly limits the performance of this error correction approach, even in the case of extremely high (or infinite) noise bias. Before this numerical study, we will briefly discuss in the next subsection how the elementary gate set $\mathcal{G}=\{\text{C}Z,\mathcal{P}_{\ket{+}},\mathcal{M}_X\}$ can be complemented by magic state preparations to ensure fault-tolerant and universal quantum computation.

\textit{Shor-style approach.}---This second approach presented in~\cite{vuillot2026} can also be studied in the framework of the architectures based on QND readout of multi-qubit Pauli \(Z\) operators, as discussed in Section~\ref{sec:ZZZ meas}. 
However, since the bias-preserving C\(Z\) gates studied in this section can also be used to perform such QND measurements, we first present this construction here. We show that, also using the bias-preserving gate set $ \cG=\{\text{C}Z,\cP_{\ket{+}},\cM_X\}$, this is a better alternative to the previous Knill-style error correction approach.

\begin{figure}[h!]
\centering
\hspace*{0.45cm} 
\begin{quantikz}[row sep=0.5em, column sep=0.35cm]
\lstick{\hspace{-2em}$\ket{+}$}&&\permute{2,1,4,3}&\midstick[4,brackets=none]{=}&\wireoverride{}\midstick[1,brackets=none]{\hspace{-5em}$\ket{+}$\hspace{-3em}}         & \gate[4, disable auto height][1][1]{\begin{matrix}
	                   M_{Z^{\otimes 4}}\\ 	\downarrow\\s
	               \end{matrix}}&\gate{Z^{m_1}}& \\
	               \lstick{\hspace{-2em}}&\gate[2, disable auto height]{\begin{matrix}
	                   M_{X^{\otimes 2}}\\ 	\downarrow\\m
	               \end{matrix}} &             &                             &\wireoverride{}\midstick[1,brackets=none]{\hspace{-5em}}         &                   & \meterD{M_X}\rstick{$\rightarrow m_1$}\\
                  \lstick{\hspace{-2em}}&        &               &                             &\wireoverride{}\midstick[1,brackets=none]{\hspace{-5em}}&                                      & \meterD{M_X}\rstick{$\rightarrow m_2$}\\
\lstick{\hspace{-2em}$\ket{+}$}&        &               &                             &\wireoverride{}\midstick[1,brackets=none]{\hspace{-5em}$\ket{+}$\hspace{-3em}}&                                      &\gate{X^sZ^{m_2}}&\\
    \end{quantikz}
    \caption{Circuit to implement an \(X^{\otimes 2}\) parity check using a \(Z^{\otimes 4}\) parity check. The measurement outcome is given by \(m=m_1\oplus m_2\). }
    \label{fig:MXX}
\end{figure}
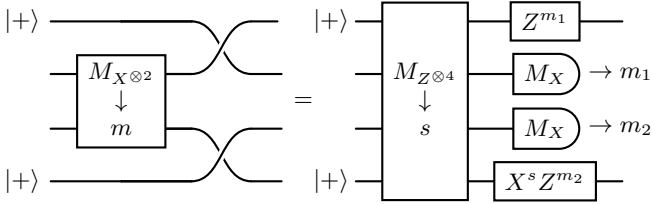

The main idea here is to use the equivalence of the logical circuits shown in Fig.~\ref{fig:MXX}. The left circuit represents the QND readout of the weight-2 $X$ stabilizer between the two middle qubits, followed by a teleportation of these two qubit states to the other two. The right circuit represents the QND readout of a weight-4 $Z$ operator followed by a destructive readout of the Pauli $X$ on the middle qubits and finally feed-forwarded Pauli  $Z$ or $X$ operations on the other qubits, based on these measurement results. 

This equivalence enables the measurement of $X$ stabilizers in a Shor-style error correction without C\(X\) gates. Several remarks are in order. In addition to the QND weight-4 $Z$ measurement that can be realized with the gate set $ \cG=\{\text{C}Z,\cP_{\ket{+}},\cM_X\}$, the circuit of Fig.~\ref{fig:MXX} also requires the application of Pauli $X$ and $Z$ gates. We note that, in practice, we can skip these Pauli gates by keeping track of them in the Pauli frame. Note, however, that an error in the QND readout of the weight-4 $Z$ operator leads to an $X$ error in the Pauli frame, or equivalently a bit-flip type error. Therefore, the bias-preserving aspect of such a $X$ stabilizer measurement relies on the fidelity of the weight-4 $Z$ operator readout. Achieve ultra-high fidelities for this readout is thus needed. Finally, we also note that the data qubits in this protocol are teleported at each step.

\begin{figure}
 \centering
    \includegraphics[width=\columnwidth]{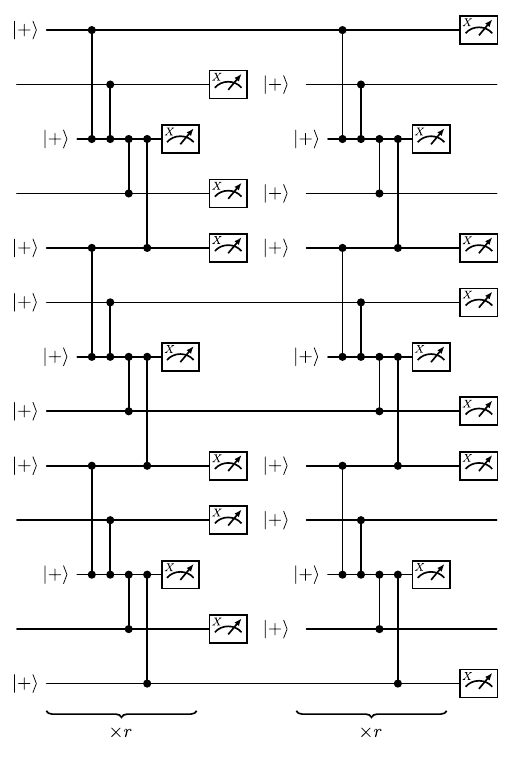}
    \caption{Shor-style error correction circuit with bias-preserving C\(Z\) gates as presented in~\cite{vuillot2026}. Here each line represents a physical qubit. At each step, by applying four bias-preserving C\(Z\) gates, we map the weight-4 $Z$ parities on the state of an ancilla qubit that is then measured in the $X$ basis. Noting that we need ultra-high fidelities for this readout, we repeat this step $r$ times, before measuring two out of four qubits in the $X$ basis as required by Fig.~\ref{fig:MXX}. At the end of this error syndrome measurement cycle, the data qubits are teleported to ancilla ones and in the next round, we therefore initialize the previous data qubits in the $\ket{+}$ states and we perform the same protocol with the new data qubits.}
    \label{fig:Vuillot-circuit}
\end{figure}

The general protocol for performing Shor-style error correction for a repetition code using this gadget with the elementary gate set $ \cG=\{\text{C}Z,\cP_{\ket{+}},\cM_X\}$ is presented in Fig.~\ref{fig:Vuillot-circuit}. The QND readout of weight-4 $Z$ stabilizers through bias-preserving C\(Z\) gates is repeated $r$ times to ensure ultra-high fidelities. The performance of this protocol is studied through the numerical simulations of Subsection~\ref{ssec:CZnumerics} and compared to the previous protocol of Fig.~\ref{fig:Aliferis-circuit}. 

\subsection{Fault-tolerant computation}\label{ssec:CZ_FT}

\begin{figure}[h!]
    \centering
    \includegraphics[width=\columnwidth]{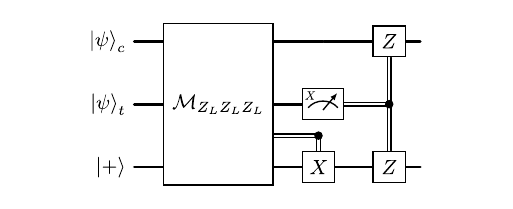}
    \caption{Fault-tolerant logical C\(X\) gate. Here, each line represents a logical qubit encoded in a code block. The central ingredient is again the fault-tolerant QND readout of a high-weight \(Z\) operator corresponding to the product of logical \(Z\) operators on three code blocks.}
    \label{fig:Aliferis CNOT}
\end{figure}

The quantum non-demolition high-weight $Z$ measurement, discussed in the Knill-style approach of the previous subsection, can also be used to implement a fault-tolerant logical C\(X\) gate between two logical qubits encoded in two code blocks, as shown in Figure~\ref{fig:Aliferis CNOT}.  We furthermore note that, with this QND high-weight $Z$ readout, we have actually engineered a fault-tolerant gate set $\tilde{\mathcal{G}}=\{\text{C}X_L,\mathcal{P}_{\ket 0_L},\mathcal{P}_{\ket +_L},\mathcal{M}_{Z_L},\mathcal{M}_{X_L}\}$ at the logical level of a repetition code or the asymmetric Bacon-Shor code. 

Combined with faulty physical preparation of the states $\ket{+i}=(\ket{0}+i\ket{1})/\sqrt 2$ and $\ket{T}=(\ket{0}+e^{i\pi/4}\ket 1)/\sqrt2$  and the use of magic state distillation, we obtain a  universal set of fault-tolerant gates. Such faulty states prepared on a single physical qubit can be teleported to the repetition code or the asymmetric Bacon-Shor code, using the same one-bit teleportation protocol (Fig.~\ref{fig:Aliferis Knill}). Indeed, one can adapt this teleportation protocol between a physical and logical qubit, using a high-weight $Z$ readout ensuring the non-destructive measurement of $Z\otimes Z_L$, followed by a destructive readout of $X$ on the physical qubit. The  injected faulty $\ket{+i}$ and $\ket{T}$ states can then be distilled following a standard approach~\cite{bravyi2005universal}, using a $[[15,1,3]]$ Reed-Muller distillation code for the state $\ket{T}$, and a $[[7,1,3]]$ Steane distillation code for the state $\ket{+i}$. For the Shor-style approach, the implementation of a universal set of fault-tolerant gates is discussed in Section~\ref{sec:ZZZ meas}.

Furthermore, the reference~\cite{webster2015reducing}  proposed a magic state protocol tailored for  biased noise qubits based on the same tool set. Assuming that the rate of correlated \(Z\) errors induced by the C\(Z\) gates is negligible, this protocol provides a  distillation  with a reduced overhead with respect to the standard approach, but only allowing to reach a logical error scaling as $\mathcal{O}(p_Z^2)$. Note, however, that all these protocols still suffer from the complexity and overhead of the high-weight \(Z\) operator readout as discussed in the next subsection.

\subsection{Numerical analysis}\label{ssec:CZnumerics}

In this subsection, we evaluate the performance of the two C$Z$-based schemes presented in~\ref{ssec:knill}. As the performance of both the logical CNOT gate and the universal gate set directly stems from that of the error correction gadgets in the Knill-style and Shor-style schemes, we focus solely on these gadgets and simulate the logical error rate of a quantum memory. We provide numerical simulations of the Knill-style error correction of Fig.~\ref{fig:Aliferis-circuit} and of the Shor-style error correction presented in Fig.~\ref{fig:Vuillot-circuit}. 

We focus on the simulations at infinite bias and as such we use the uniform phase-flip circuit-level noise model described in Table~\ref{tab:aliferis-noise-model} indexed by the total error rate per operation $p_Z$. This model is chosen so that for all operations, i.e. idle, measurement, state preparation, and two-qubit gate, the error probability per operation is equal to $p_Z$. We note that we do not consider correlated phase-flip noise on the C$Z$ gate as the $Z$ operator commutes with the gate and such errors are absent in a number of physical implementations (note that simulations with correlated phase-flip errors do not represent a significant difference~\cite{vuillot2026}).

\begin{table}[htbp]
\renewcommand{\arraystretch}{1.6} 
\begin{ruledtabular}
\begin{tabular}{lcc}
Operation & Error type & Probability \\
\colrule
Idle      & $Z$    & $p_Z$ \\
\colrule
Prep. $\cP_{\ket{+}}$  & $Z$    & $p_Z$ \\
\colrule
Meas. $\cM_{X}$  & assignment error    & $p_Z$ \\
\colrule
Two-qubit C\(Z\)          
                       & $ZI, IZ$                             & $\frac{p_Z}{2}$ \\
\end{tabular}
\end{ruledtabular}
\caption{\label{tab:aliferis-noise-model} 
Error model for the error correction circuits of Fig.~\ref{fig:Aliferis-circuit} and Fig.~\ref{fig:Vuillot-circuit}. The total error probability per operation is $p_Z$, and the noise bias is infinite (no bit-flip type errors).}
\end{table}

\begin{figure}[h!]
    \centering
    \includegraphics[width=\columnwidth]{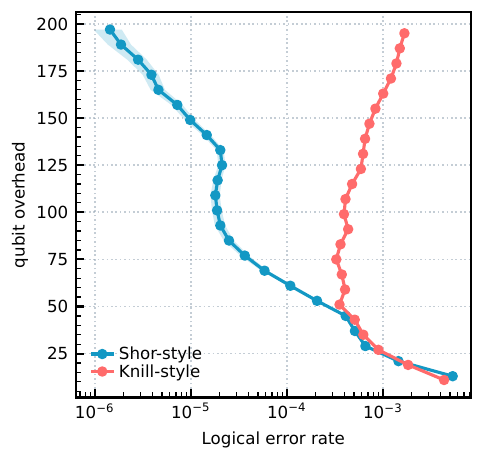}
    \caption{Qubit overhead as a function of the targeted logical error rate. We consider a physical phase-flip error rate of $p_Z=10^{-3}$ and an infinite noise bias, i.e. $p_X=p_Y=0$. This plot illustrates the achievable logical error rate with a repetition code while increasing the distance of the repetition code. The different dots correspond to different distances $d$. At each distance, the qubit overhead corresponds to the optimal $r_1$ and $r_2$ or $r$ providing the best logical error rate. The logical error rate plateau in the Shor-style scheme arises from the discrete nature of the number of repetitions of the multi-$Z$ measurement $r$. We note that, in the Knill-style error correction, for large distances, the logical error rate increases. This is because, at large code distances, preparing the $\ket{\text{GHZ}}$ state incurs a substantial time overhead, allowing too many phase-flip errors to accumulate for the repetition code to correct them. This results in an optimal finite distance to minimize the logical error rate.}
    \label{fig:CZ_overhead}
\end{figure}

In Figure~\ref{fig:CZ_overhead}, we compare the qubit overhead required to reach a given logical error rate at a physical error rate of $p_Z = 10^{-3}$. We simulated the circuit over $d$ rounds of $X$-stabilizer measurements, where $d$ is the distance of the repetition code, using the Stim Clifford simulator~\cite{gidney2021stim} and decoded the syndromes with the PyMatching decoder~\cite{higgott2025sparse}. The logical error rate is reported per round of $X$-stabilizer measurements, and the optimal $r_1$ and $r_2$ or $r$ that minimizes the logical error rate for each qubit count is selected. For both schemes, the quantum information support oscillates between physical qubits, and all the qubits, including ancilla qubits, are included in the qubit overhead.

The Knill-style error correction performs really poorly at $p_Z=10^{-3}$. This can be explained by the fact that the time overhead required to extract the $X$ stabilizers of the repetition code, equal to $r_1 \times r_2$ rounds of physical $ZZ$ measurements, is very long. During this long extraction time, the data qubits accumulate phase-flip errors. This eventually leads to a logical error if more than half of the data qubits undergo an error.

In contrast, the Shor-style error correction has a more favorable scaling. This is explained by the fact that multi-$Z$ measurements are performed at the physical level and do not require the preparation of large $\GHZ$ states. This implies that the number of repetitions $r$ is smaller than $r_1 \times r_2$, resulting in a much better logical error rate.

However, achieving algorithmically relevant logical error rates with the Shor-style protocol at $p_Z=10^{-3}$ still requires a massive overhead that is not competitive with a surface code approach with standard qubits. For comparison, with a depolarizing error probability of $p=10^{-3}$ per operation, a distance-9 surface code would suffice to reach a logical error rate of $10^{-6}$ while only requiring 161 qubits~\cite{fowler2018low}. These observations hint at the fact that we cannot benefit from the noise bias in a practical regime when the set of elementary physical operations that preserve the noise bias is limited to $\{\text{C}Z,\mathcal{P}_{\ket{+}},\mathcal{M}_X\}$. In other words, with this restrictive set of bias-preserving operations, it is usually better to forget about the noise bias and rely on standard error correction approaches designed for depolarizing noise. 

\section{\label{sec:CNOT}C\(X\)-based architecture}

As seen in Section~\ref{sec:CZ}, the noise bias, combined with a bias-preserving $\text CZ$ gate at a typical phase-flip rate of $p_Z=10^{-3}$, cannot reduce the overhead compared to a standard approach with a depolarizing noise model of similar rate. This is due to the very restrictive set of bias-preserving physical gates given by  $\{\text CZ, \mathcal{P}_{\ket +},\mathcal{M}_X\}$. Notably, the absence of a bias-preserving C\(X\) gate renders the syndrome extraction complicated, and explains the limited performance of $\text C Z$-based biased-noise qubit architectures. However, the C\(X\) gate does not map phase-flip errors to bit-flip errors and could, in theory, be bias-preserving. Indeed, as discussed in Section~\ref{sec:bias}, the bosonic cat qubits or the large nuclear spin cat qubits, in addition to their large noise bias, support the implementation of a bias-preserving C\(X\).  In this section, we review a set of results that show that the bias-preserving C\(X\) unleashes the real power of biased noise qubits. 

In the first Subsection~\ref{sec:qec noise bias}, we review various quantum error correction strategies tailored to benefit from the noise bias, with an elementary bias-preserving gate set $\{\text C X, \text C Z, \text C Y,\mathcal{P}_{\ket +},\mathcal{P}_{\ket 0},\mathcal{M}_X,\mathcal{M}_Z\}$. We will provide a detailed analysis of their performances and a comparison of their overhead when considering different regimes of noise bias in Subsection~\ref{sec:code_comparison}. Then in subsection~\ref{ssec:cx_ft}, we will show that the implementation of a bias-preserving C\(X\) also leads to huge savings with customized fault-tolerant protocols for logical gates. Finally, in the case of non-Clifford gates, we show that the most economical fault-tolerant construction relies on the bias-non-preserving $X^{\pm1/4}$ gate. We also discuss a construction based on bias-preserving CC\(X\) gates. The references~\cite{guillaud2019repetition} and~\cite{albert2016holonomic} provide detailed descriptions of the implementation of these elementary physical operations. We only note here that they represent different levels of complexity from an experimental point of view, with the bias-preserving CC\(X\) requiring the more complex engineering of high-order nonlinear interactions and the bias-non-preserving $X^{\pm1/4}$ merely requiring modulation of microwave drives.

\subsection{Biased-noise error correction}
\label{sec:qec noise bias}

The noise bias property and the implementation of a bias-preserving C\(X\) gate can greatly reduce the overhead required for QEC. In this subsection, we present various  error correction protocols  tailored for biased-noise qubits that possess a bias-preserving implementation of C\(X\). We  compare the performance of these schemes in various noise regimes. We recall that a code encoding $k$ logical qubits in $n$ physical qubits with distance $d$ is denoted as a $[n,k,d]$-code for a classical code correcting one type of error and as $[[n,k,d]]$ for a quantum code correcting both bit-flip and phase-flip errors.

\subsubsection{Phase-flip error correction}\label{ssec:phase-flip}

For some biased-noise qubits, such as dissipative cat qubits, the bit-flip errors are sufficiently rare such that certain quantum algorithms can be run without the requirement of further bit-flip error correction, as seen in Subsection~\ref{ssec:engineered}. In these extreme noise bias regimes, we therefore solely need to correct phase-flip errors.

\begin{figure}
    \centering
    \includegraphics[width=\columnwidth]{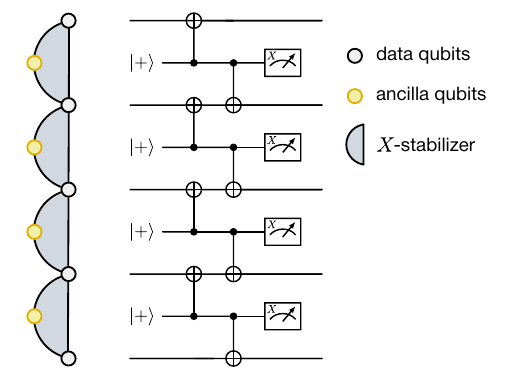}
    \caption{Repeated stabilizer measurements of a $d=5$ phase-flip repetition code.}
    \label{fig:repetition_code}
\end{figure}

\textit{Repetition code.}---The phase-flip error correction can be done through a phase-flip repetition code~\cite{guillaud2019repetition,chamberland2022building}. The experimental simplicity of such a code, requiring low connectivity in a 1D layout, makes this approach very useful for early demonstrations~\cite{putterman2025hardware}. The stabilizers of the repetition code are given by $\{ X_i X_{i+1} \}_{1 \le i \le d-1}$ and the logical operators by $Z_L = Z_1\otimes Z_2\otimes\cdots\otimes Z_d$ and $X_L = X_k$, for any $k\in\{1,\cdots, d\}$. The code has a total of $2d-1$ qubits and the stabilizers can be measured repeatedly with the circuit illustrated in Figure~\ref{fig:repetition_code}.

Note that a physical bit-flip error at any location within the circuit (except trivial locations on the ancilla qubits) creates a logical bit-flip error. Thus, quantum computation with repetition cat codes requires the cat qubits to experience no bit-flip errors during the whole algorithm. With their present bit-flip rates~\cite{reglade2024quantum,rousseau2025enhancing}, dissipative cat qubits concatenated with a repetition code may be capable of supporting the first useful quantum algorithms~\cite{beverland2022assessing}. The key experimental challenge remains to demonstrate C\(X\) gates, that not only maintain the same level of bit-flip error probability, but also admit low enough phase-flip error rates, below the error correction threshold of the repetition code. Although the repetition code has a circuit-level noise threshold  of about $4\%$~\cite{ruiz2025ldpc}, well above that of the surface code around $0.8\%$~\cite{higgott2023improved}, it could still be quite challenging to demonstrate bias-preserving C\(X\) gates with phase-flip error probability below this value. Indeed, as discussed in Section~\ref{sec:bias}, while the fidelity of a fast C\(X\) gate is limited by non-adiabatic effects, a slow one suffers from the undesired single photon loss of the cat mode. Reaching high-fidelity bias-preserving C\(X\) gates thus requires a strong separation of time-scales between the engineered two-photon dissipation  and the undesired single-photon loss. Alternatively, a cat-transmon hybrid scheme was recently developed  and operated below the phase-flip repetition code threshold~\cite{hann2025hybrid,putterman2025hardware}. In this approach, the data cat qubits are coupled to transmon measurement qubits and a bias-preserving C\(X\) is applied between the transmons and the cat qubits.  While high-fidelity C\(X\) gates are simpler to achieve in this architecture, the bit-flip suppression is only limited to first order with respect to the transmon's decay channel.

If the repetition code is attractive for its experimental simplicity, as an $[n,1,n]$-code, it also suffers from a low encoding rate of $k/n = 1/d$. For typical distances in the range of $d=10$--$20$, 20 to 40 physical qubits (counting the ancilla qubits) are required to encode one logical qubit.

\textit{Cellular automaton codes.}---Quantum low-density parity-check (qLDPC) codes are a class of quantum error-correcting codes characterized by the fact that each data qubit participates in only a constant number of checks, and each check involves only a constant number of data qubits. They have the ability to encode quantum information at higher rates, that is, using fewer physical qubits per logical qubit. The key insight is that, instead of encoding logical qubits separately in different blocks, qLDPC codes encode multiple logical qubits within a single block, allowing them to share physical qubits and thereby reducing resource requirements. After breakthrough theoretical results demonstrating the existence of ``good'' qLDPC codes~\cite{panteleev2022asymptotically,leverrier2023decoding}, i.e. codes where both the number of logical qubits and the distance scale, asymptotically, linearly with the number of physical qubits, several small and practical constructions were identified~\cite{bravyi2024high}. However, these constructions rely on long-range interactions between qubits, which are challenging for platforms such as superconducting circuits. These long-range interactions are nonetheless needed, as quantum codes restricted to short-range interaction in a 2D layout must obey the Bravyi-Poulin-Terhal (BPT) bound~\cite{bravyi2010tradeoffs}, i.e., $k \le c n/d^2$, where $k$ is the number of logical qubits, $n$ the number of physical qubits, and $c$ a constant. Note that this bound is saturated by the surface code with $k=1$ and $d=\sqrt{n}$. 

\begin{figure}
    \centering
    \includegraphics[width=\columnwidth]{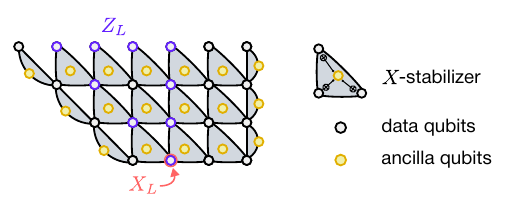}
    \caption{Cellular automaton code of distance $d_Z = 9$ and $d_X = 1$ encoding 4 logical qubits. The weight-2 and weight-3 stabilizers of the code can be measured on a rectangular hexagonal lattice of qubits. In cellular automaton codes, the number of logical qubits is determined by the number of physical data qubits in the bottom row. In this example, $k=4$. The data qubits highlighted in blue are the support of one of the minimal weight $Z_L$ operators, while the corresponding $X_L$ operator (in red) is supported on the qubit on the bottom line. Cellular automaton codes allow for a higher encoding rate than the repetition code, as for instance here $k/n = 4/22 \approx 0.18$ whereas a $d_Z=9$ repetition code has a rate of $k/n = 1/9 \approx 0.11$. As the distance increases (by increasing the height of the lattice), the rate improvement increases over the repetition code.}
    \label{fig:cellular_automaton}
\end{figure}

Biased-noise qubits, if they possess negligible bit-flip error rate, allow us to  circumvent this bound as they require codes that only correct phase-flip errors~\cite{ruiz2025ldpc}. These codes have a looser bound $k \le c n/\sqrt{d}$~\cite{bravyi2010tradeoffs}, allowing for higher-rate codes than the repetition code (where $k=1$ and $d=n$), while staying with a 2D local layout. Specifically, the family of cellular automaton codes~\cite{bravyi2010tradeoffs,yoshida2013information,nixon2021correcting,chowdhury1994design,newman1999glassy} offers high-rate constructions while remaining relatively simple to implement. Figure~\ref{fig:cellular_automaton} illustrates the simplest cellular automaton code, which can be implemented on a 2D grid of qubits.

\begin{figure}
    \centering
    \includegraphics[width=\columnwidth]{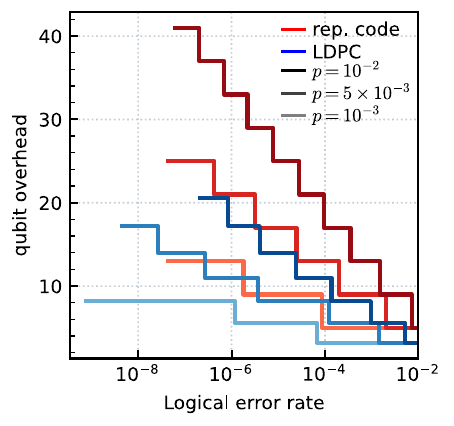}
    \caption{Logical error rate of a phase-flip cellular automaton code after $d$ rounds of QEC. A circuit-level noise model with a noise bias of $\eta = + \infty$ is used.}
    \label{fig:numerics cellular automaton}
\end{figure}

\begin{table}
\renewcommand{\arraystretch}{1.6} 
\begin{ruledtabular}
\begin{tabular}{lcc}
Operation & Error type & Probability \\
\colrule
Idle      & $Z$    & $p_Z$ \\
\colrule
Prep. $\cP_{\ket{+}}$  & $Z$    & $p_Z$ \\
\colrule
Meas. $\cM_{X}$  & assignment error    & $p_Z$ \\
\colrule
Two-qubit C\(X\)          
                       & $ZI, IZ, ZZ$                             & $\frac{p_Z}{3}$ \\
\end{tabular}
\end{ruledtabular}
\caption{\label{tab:ldpc-noise-model} 
Error model for the error correction circuits of Fig.~\ref{fig:repetition code} and Fig.~\ref{fig:cellular_automaton}. The total error probability per operation is $p_Z$, and the noise bias is infinite (no bit-flip type errors).}
\end{table}

Figure~\ref{fig:numerics cellular automaton} compares the overhead needed to reach a given logical error rate for the cellular automaton code of Figure~\ref{fig:cellular_automaton} and for the repetition code. The noise model used is presented in Table~\ref{tab:ldpc-noise-model}. We simulated the circuit over $d$ rounds of $X$-stabilizer measurements using the Stim Clifford simulator~\cite{gidney2021stim} and decoded the syndromes with the PyMatching decoder~\cite{higgott2025sparse}. The logical error rate per round of stabilizer measurement is reported. As expected, thanks to its increased rate, the cellular automaton codes allow us to reduce the overhead. Although their threshold is slightly lower than that of the repetition code, their reduced overhead and experimental simplicity make them strong candidates for highly biased qubits. In a recent result~\cite{ruiz2025ldpc}, we numerically optimized such classical LDPC codes to find the best encoding rate for algorithmically relevant code distances. In particular, we find a family of codes with parameters $[n=165+8l,k=34+2l,d=22]$, which can encode 100 logical qubits with a total logical error probability below $10^{-8}$ on a 758 cat-qubit chip with physical phase-flip probability of $10^{-3}$ per qubit and per operation. This is a qubit overhead of less than 8 per logical qubit, noting that in this reference, we had considered a different noise model roughly corresponding to an error rate in the order of $10^{-3}$ for single-qubit operations and of $10^{-2}$ for  C\(X\) operation. Comparing to the similar simulations of Fig.~\ref{fig:CZ_overhead} for the C\(Z\)-based architecture (where with $p_Z=10^{-3}$ for all operations, we require at best an overhead of 200 to reach a logical error rate of $10^{-6}$), we observe the drastic change of situation with bias-preserving C\(X\). 

\subsubsection{Asymmetric quantum error correction}
\label{sec:quantum_codes}

For more moderate noise bias regimes, or even with strong noise biases in the case of large-scale quantum algorithms, one might need to ensure some level of bit-flip error correction. In this subsection, we review the most efficient schemes that have been proposed in the literature and provide a detailed comparison. We will discuss how tailoring quantum error correction codes to the noise bias significantly reduces the qubit overhead, either by utilizing asymmetric distances or by increasing the fault-tolerance threshold.

\begin{figure}
    \centering
    \includegraphics[width=\columnwidth]{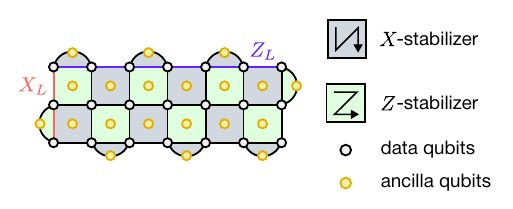}
    \caption{A rectangular surface code of distance $d_X = 3$ and $d_Z = 7$. The arrows on the stabilizers represent the order in which the CNOT gates have to be executed to not reduce the distance of the code while measuring the stabilizers.}
    \label{fig:thin surface code}
\end{figure}

\textit{Thin surface code.}---Rectangular or thin surface codes~\citep{chamberland2022building} can be used with a  bit-flip distance $d_X$ smaller than the phase-flip one $d_Z$ as illustrated in Figure~\ref{fig:thin surface code}. Note that this strategy still requires bias-preserving C\(X\) gates for syndrome extractions. Indeed, the benefit of asymmetric distance is lost if such syndrome extractions depolarize the noise. Although this approach increases the connectivity from two to four compared to repetition codes, thin surface codes can be still naturally implemented on a two-dimensional grid of qubits. Compared to the repetition code, in addition to a larger code size by a factor of $d_X$, the surface code also admits a lower phase-flip threshold. For instance, for a $d_X=3$ surface code, the phase-flip threshold is less than half of that of the repetition code (cf.~\cite{ruiz2025ldpc} vs.~\cite{hann2025hybrid}). However, it still represents a reduction of the overhead with respect to a symmetric surface code, as soon as the noise bias $\eta > 1$.

\begin{figure}
    \centering
    \includegraphics[width=\columnwidth]{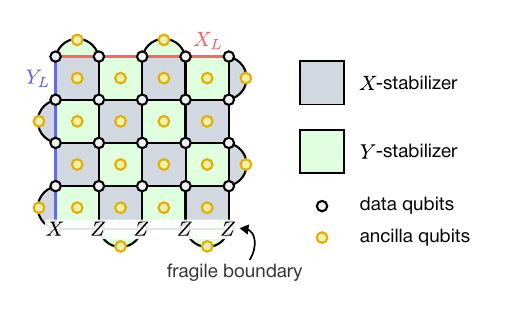}
    \caption{XY surface code. A fragile spatial boundary is highlighted. The $Y$-type stabilizers can be measured with C$Y$ gates or alternatively compiled with CNOT and $S$ gates. The order of the gates must be the same as that of the surface code presented in Figure~\ref{fig:thin surface code}.}
    \label{fig:XY code}
\end{figure}

\textit{XY code.}---Calderbank-Shor-Steane (CSS) codes~\cite{calderbank1996good,steane1996multiple} admit stabilizers that are products of either only Pauli $X$  operators or only Pauli $Z$ ones. However, for biased noise qubits, it could be relevant to use non-CSS codes. Two families of codes have been widely studied in this context, which we describe in the following. The XY code consists in replacing the $Z$ stabilizers of the surface code by $Y$ ones while the XZZX code relies on mixed stabilizers. Although the thin surface code is designed to have a smaller distance against the rarer error, none of the $Z$ stabilizers detect the dominant phase-flip errors. In contrast, by simply replacing the $Z$ stabilizers of the surface code with $Y$ stabilizers~\cite{tuckett2018ultrahigh} as shown in Figure~\ref{fig:XY code}, these stabilizers can both detect bit-flip errors and phase-flip ones. While the code admits a distance $d$, in the absence of bit-flip errors, the phase-flip distance is equal to $d^2$, making the code very robust to phase-flip errors in the limit of large noise bias.

This property is however very fragile as a single bit-flip error is enough to reduce the phase-flip distance to $d$. This phenomenon known as \textit{fragile spatial boundaries}~\cite{higgott2023improved} is illustrated in Figure~\ref{fig:XY code}. This clearly also forbids the rectangular versions of the XY code as, in the presence of single bit-flip type error, the phase-flip distance would be reduced to the shorter edge of the surface code.  In a similar manner, when a logical measurement is performed on the code, only half of the stabilizers, either the $X$ or $Y$ stabilizers, can be reconstructed from the data qubit measurements. As a result, the effective distance against phase-flip errors is reduced to $d$. This effect, known as the \textit{fragile temporal boundary}~\cite{higgott2023improved}, also arises when the code is initialized in the $X$ or $Y$ basis. More importantly, this problem also arises during lattice surgery operations required for fault-tolerant gates, making all two-qubit logical operations on the XY code far noisier than idling.

The XY code also poses additional challenges for decoding. $Z$ errors on data qubits trigger four stabilizers, preventing the use of traditional matching decoders. To address this decoding challenge, alternatives such as the belief-matching decoder have been proposed. This approach uses iterations of belief propagation to adjust the edge weights for a subsequent round of standard matching decoding~\cite{higgott2023improved}. While practical, this decoder remains far from optimal. Furthermore, regardless of the decoding strategy, the logical error rate is fundamentally constrained by weak temporal and spatial fragile boundaries. Nevertheless, when optimally decoded, the XY code exhibits an increased threshold close to the one of the repetition code~\cite{tuckett2018ultrahigh}, making it a compelling alternative to the surface code for early, small-scale experiments with biased-noise qubits.

\begin{figure}
    \centering
    \includegraphics[width=\columnwidth]{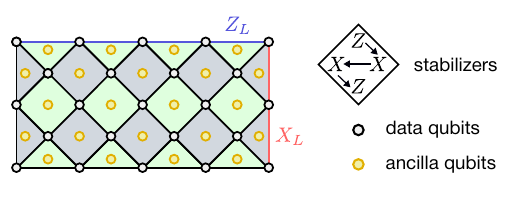}
    \caption{XZZX surface code. The stabilizers can be measured with C\(X\) and C$Z$ gates. The arrows on the stabilizers represent the order in which the gates have to be executed to not reduce the distance of the code while measuring the stabilizers.}
    \label{fig:XZZX code}
\end{figure}

\textit{XZZX code.}---The XZZX surface code is a recently proposed variant of the surface code designed to take advantage of the noise bias. Both $X$ and $Z$ stabilizers are here replaced by the unique $XZZX$-type stabilizer~\cite{bonilla2021xzzx}, as shown in Figure~\ref{fig:XZZX code}. In the absence of $X$-type errors, the XZZX code is equivalent to $2 d_X-1$ independent strips of phase-flip repetition code, where similarly to the XY code, every stabilizer is useful to detect phase-flip errors. Note in particular that this time, similarly to the standard surface code, we can choose asymmetric distances to further reduce the hardware overhead of QEC. It was shown that in the limit of infinite noise bias, the XZZX code has the same threshold as the repetition code~\cite{bonilla2021xzzx}. However, as it needs to be implemented on an unrotated lattice to preserve the asymmetric distances $d_X$ and $d_Z$, its qubit overhead is twice that of the surface code for given distances $d_X$ and $d_Z$.

\textit{Hierarchical error correction with code concatenation.}---Code concatenation has recently emerged as a promising approach, even for platforms that are constrained to locality in 2D~\cite{pattison2025hierarchical,gidney2025yoked}. In these constructions, the surface code is used at the first level because of its high threshold, but once a sufficiently low logical error rate is achieved, this first layer is concatenated with a higher-rate code compiled at the logical level, like one would do for an algorithm. In Ref.~\cite{gidney2025yoked}, the authors concatenated surface codes with  $[[n,n-2,2]]$ and $[[n,n-4\sqrt{n},4]]$ high-rate codes, showing that to reach a logical error rate of $10^{-12}$ at a physical error rate of $p=10^{-3}$, the number of qubits could be divided by 3 compared to a standard surface code approach. To stay with a 2D local architecture, the checks of the outer code are measured with lattice surgery operations between the inner patches of surface codes. To prevent the routing qubits required for lattice surgery from dominating the overhead, the authors choose to measure the checks sequentially.

\begin{figure}[h!]
    \centering
    \includegraphics[width=\columnwidth]{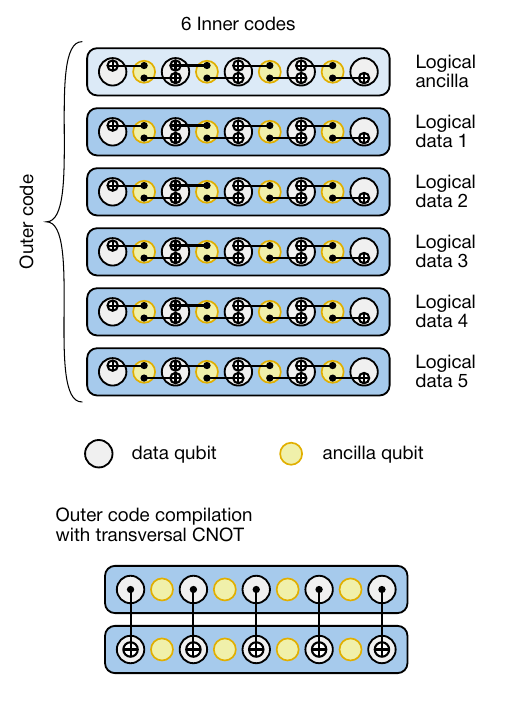}
    \caption{Layout of concatenated repetition codes. The inner codes are repetition codes and the outer code is a $n = 5$ qubit code. The 6 blocks of separate inner codes are the input data and ancilla qubits of the outer code. Logical CNOT and SWAP gates are executed by transversal CNOT gates to implement the syndrome extraction of the outer code using the ancilla logical qubit. Depending on whether we want to optimize the space or the time overhead, either a single logical ancilla that measures the checks sequentially can be used or more logical ancilla can be included to measure the checks in parallel.}
    \label{fig:elevator_codes}
\end{figure}

A similar approach can be used for biased-noise qubits, but the construction can be simplified by concatenating two classical codes, a phase-flip repetition code at the first level, and a high-rate bit-flip code at the second level~\cite{shanahan2026elevator}. The measurements of the outer code checks can also be simplified compared to lattice surgery that requires both a large space and time overhead. Indeed, repetition codes allow for the implementation of a transversal CNOT gate while staying with a 2D architecture. Thus, the repetition codes can be stacked vertically, and the outer code checks measured via transversal CNOT gates, as shown in Figure~\ref{fig:elevator_codes}. More precisely, a repetition code acting as a logical ancilla is prepared in $\ket{0}_L$ and then swept through the vertical stack, performing either a logical CNOT or SWAP gate, depending on the outer-code check to be measured. Measuring the ancilla in the $Z$ basis then yields the error syndrome. If the outer code is small enough, this strategy enables the fast measurement of the outer code checks at a low qubit overhead cost. In the construction that was nicknamed \textit{elevator codes}~\cite{shanahan2026elevator}, due to the movement of the logical ancilla, the outer bit-flip codes of parameters $[15, 9, 3]$, $[15, 6, 5]$, and $[16, 3, 8]$ were evaluated.

\subsection{Code comparison}
\label{sec:code_comparison}

\begin{table}
\renewcommand{\arraystretch}{1.6} 
\begin{ruledtabular}
\begin{tabular}{lcc}
Operation & Error type & Probability \\
\colrule
\multirow{2}{*}{Idle} & $Z$    & $p_Z$ \\
                      & $X, Y$ & $\frac{p_X}{2}$ \\
\colrule
Prep. $\mathcal{P}_{|+\rangle}$  & $Z$    & $p_Z$ \\
\colrule
Prep. $\mathcal{P}_{|0\rangle}$  & $X$    & $p_X$ \\
\colrule
Meas. $\mathcal{M}_{X}$  & assignment error    & $p_Z$ \\
\colrule
Meas. $\mathcal{M}_{Z}$  & assignment error    & $p_X$ \\
\colrule
\multirow{2}{*}{Two-qubit C\(X\)}         
                       & $ZI, IZ, ZZ$                             & $\frac{p_Z}{3}$ \\
                       & $P_1 P_2$, $P_1 \text{ or } P_2 \in \{X,Y\}$ & $\frac{p_X}{12}$ \\
\end{tabular}
\end{ruledtabular}
\caption{\label{tab:quantum-code-noise-model} 
Error model for the error correction codes presented in Subsection~\ref{sec:quantum_codes}.  The noise bias is defined as $\eta = p_Z/p_X$.}
\end{table}

\begin{figure*}
    \centering
    \includegraphics[width=\textwidth]{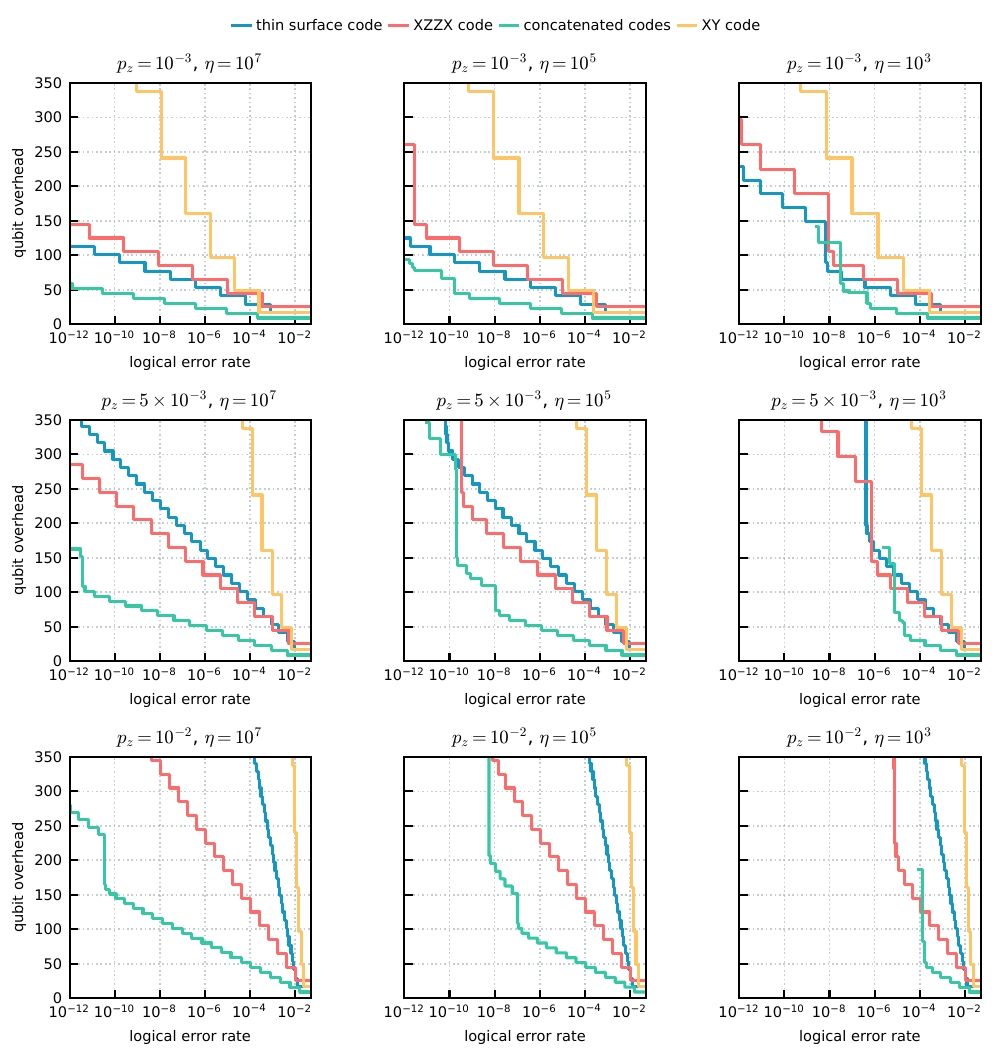}
    \caption{Qubit overhead required by different error correction strategies to reach a given logical error rate in various noise regimes. The logical error rate is the sum of the $X$ logical error rate and the $Z$ logical error rate (or the $Y$ logical error rate for the XY code), reported per round of error correction. For logical error rates that are inaccessible to Monte Carlo simulations, the results are extrapolated from higher physical error rates or lower code distances. These extrapolations are performed following the same method as in~\cite{shanahan2026elevator}.}
    \label{fig:code comparison}
\end{figure*}

In this subsection, we compare the surface code, the XY code, the XZZX code and the concatenated codes to identify which code has the smallest overhead in various noise regimes. For all codes, we use the error model presented in Table~\ref{tab:quantum-code-noise-model}. The results are presented in Figure~\ref{fig:code comparison}. We simulated each circuit over $d_Z$ rounds for the $X$ logical error rate and over $d_X$ rounds for the $Z$ logical error rate using the Clifford simulator Stim~\cite{gidney2021stim}. The PyMatching decoder~\cite{higgott2025sparse} is used to decode the thin surface code and the XZZX code, while BP+OSD is used to decode concatenated codes~\cite{panteleev2021degenerate,roffe2020decoding}, and belief-matching is used to decode the XY code~\cite{higgott2023improved}. The logical error rate per round of stabilizer measurement is reported. In the following, we discuss the main features of this comparison and explain the mechanisms responsible for the different behaviors observed across the various noise regimes.

The XZZX code and the surface code are beneficial in different error regimes. At high phase-flip error rates ($p_Z = 10^{-2}$), the XZZX code benefits from its high phase-flip threshold and is advantageous over the surface code. However, at $p_Z = 10^{-3}$, the surface code lower qubit overhead outweighs the XZZX code higher threshold, giving the surface code a slight advantage over the XZZX code. The XY code, on the other hand, is penalized by its fragile spatial boundaries and has a higher overhead than the surface code or the XZZX code. We note that, while the fragile temporal boundary does not have a large impact here, because it only affects the first and last rounds of the simulation, it would have a larger impact when performing a quantum algorithm during lattice surgery operations. We also note that the decoding is performed with belief-matching~\cite{higgott2023improved}, in order to handle the hyperedges, but it might be far from optimal.

Concatenated codes have a clear advantage in the large-bias regime when $\eta \ge 10^{5}$. The inner repetition codes provide a high threshold against phase-flip errors, resulting in a low overhead even at relatively large phase-flip error rates. At the same time, the high rate of the outer code allows bit-flip errors to be suppressed with only a modest overhead. Overall, this shows that if the separation between the phase-flip and the bit-flip error rates is large enough, they should be corrected at different levels of the code hierarchy. This approach, however, requires the physical bit-flip error rate to remain sufficiently low. Indeed, a bit-flip error on any qubit of the repetition code induces a bit-flip error at the logical level of the repetition code, which is corrected only by the outer code. Moreover, the outer-code checks are measured less frequently than the inner repetition-code checks. Thus, the physical bit-flip error rate must remain well below the outer-code threshold for this strategy to be efficient. Consequently, there exists a value of the noise bias above which concatenated codes become advantageous.

\subsection{Fault-tolerant quantum computation}\label{ssec:cx_ft}

If the noise bias provides an advantage for the quantum memory overhead, its properties can also be leveraged for fault-tolerant quantum computing. The Clifford group, defined as the set of unitaries that map Pauli operators to Pauli operators under conjugation, plays a central role in fault-tolerant quantum computation. While some 2D codes, such as the color code, can implement the entire Clifford group transversally, a series of no-go theorems establishes fundamental limitations on implementing gates beyond the Clifford group in this setting. In particular, the Bravyi--Koenig theorem shows that, in two dimensions, any encoded gate implemented by a constant-depth local circuit must belong to the Clifford group~\cite{bravyi2013classification}. To circumvent this limitation, magic states are prepared offline and consumed through gate teleportation to implement logical non-Clifford gates~\cite{zhou2000methodology}. In this section, we show that although the implementation of Clifford gates for biased-noise qubits is largely similar to that for conventional qubits, significant advantages can be obtained in the preparation of magic states.

\subsubsection{Clifford group}

\begin{figure}
    \centering
    \includegraphics[width=\columnwidth]{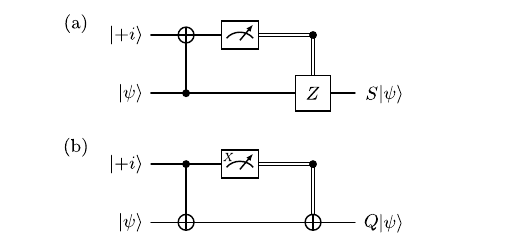}
    \caption{(a) Gate teleportation of the $S$ gate from the $\ket{+i}$ state. (b) Gate teleportation of the $Q$ gate from the $\ket{+i}$ state.}
    \label{fig:S_gate}
\end{figure}

In order to implement the full Clifford group, we focus on the implementation of the following gates $\mathcal F = \{\mathcal{P}_{\ket 0_L},\mathcal{P}_{\ket +_L},\mathcal{M}_{Z_L},\mathcal{M}_{X_L}, \text{C}X_L, \mathcal{P}_{\ket{+i}_L}\}$. The $\ket{+i}_L$ state allows us to perform the $S$ gate with the teleportation circuit shown in Figure~\ref{fig:S_gate}(a) if we are already equipped with the CNOT gate and measurement in the $Z$ basis. Note that the $Z$ gate in this circuit, like all Pauli gates during the computation, does not need to be applied physically but can be simply tracked classically by updating the Pauli frame. Similarly, the gate $Q = SHS = X^{-1/2}$ can be implemented~\cite{chamberland2022building}, as shown in Figure~\ref{fig:S_gate}(b). The Hadamard gate can then be implemented as $H = S^\dagger Q S^\dagger$. The $S^\dagger$ gate is similar to the $S$ gate up to the $Z$ Pauli gate, which can be tracked in software and does not require a different resource state than the $S$ gate. Thus, all Clifford gates can be implemented. In the following, we describe how to implement physically the logical gates of the set $\mathcal F$.

\textit{Pauli preparation and measurement.}---The preparation of the $\ket{+}_L$ and $\ket{0}_L$ states can be performed by leveraging stabilizer measurements. For all the CSS codes (the repetition code, the cellular automaton codes, the surface code and the elevator codes), preparing all physical qubits in $\ket{+}$ ($\ket{0}$) and then performing rounds of stabilizer measurement prepares the state $\ket{+}_L$ ($\ket{0}_L$). For codes encoding multiple logical qubits, such as cellular automaton codes and elevator codes, this method prepares all logical qubits in a block in the same state (or basis, up to Pauli correction). However, this method does not allow different logical qubits within the same block to be prepared simultaneously as eigenstates of different Pauli operators. For non-CSS codes, such as the XZZX and XY codes, the same techniques can be applied by preparing physical qubits in the appropriate basis: either $\ket{+}$ or $\ket{+i}$ for the XY code, and a combination of $\ket{+}$ and $\ket{0}$ for the XZZX code. Conversely, logical Pauli measurements are implemented by measuring each physical qubit in the same basis in which it is prepared for state initialization.

\begin{figure}
    \centering
    \includegraphics[width=\columnwidth]{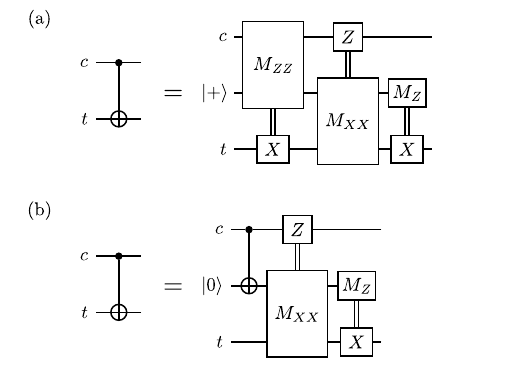}
    \caption{CNOT gate via lattice surgery. (a) The scheme requires a third logical qubit prepared in the state $\ket{+}$. The operation is realized through a sequence of $M_{ZZ}$ and $M_{XX}$ joint measurements, followed by Pauli corrections conditioned on the measurement outcomes. (b) Equivalent implementation with the logical ancilla qubit initialized in the $\ket{0}$ state. In both cases, the logical CNOT is obtained without requiring direct interaction between the control and target qubits, allowing long-range gates. In both schemes, the multi-Pauli measurements need to be repeated for several rounds (typically, either $d_Z$ or $d_X$) to ensure fault tolerance.}
    \label{fig:lattice_surgery}
\end{figure}

\textit{CNOT gate.}---To complete the Clifford group, the CNOT gate and the preparation of the $\ket{+i}$ state are required. Since the latter is closely similar to the non-Clifford magic state preparation, we postpone its discussion to the next subsection and focus here on the CNOT gate. Two challenges arise in the implementation of a logical CNOT gate. First, even if CSS codes support the transversal implementation of the logical CNOT, we must be able to perform long-range logical CNOT operations required by the algorithm between qubits that may be spatially separated, while staying with a 2D architecture. Second, for codes encoding multiple logical qubits, we must be able to address individual logical qubits within the code block without affecting the remaining logical qubits. Lattice surgery solves both of these problems by implementing CNOT gates with multi-Pauli logical measurements, as shown in Figure~\ref{fig:lattice_surgery}. As demonstrated in Ref.~\cite{gouzien2023performance} and~\cite{ruiz2025ldpc}, for the repetition code or cellular automaton code, it is preferable to use the variant of panel (b), with a transversal CNOT followed by a $XX$ measurement, while for the quantum codes, the variant of panel (a) is preferred. We note that for the cellular automaton codes, this operation requires a bi-layer architecture, which could for instance be realized on a flip chip, as detailed in Ref.~\cite{ruiz2025ldpc}.

\subsubsection{Magic states}

\begin{figure}
    \centering
    \includegraphics[width=\columnwidth]{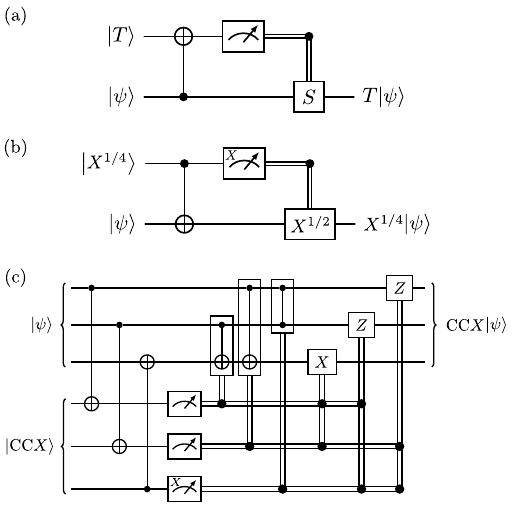}
    \caption{(a) $\ket{T}$ state teleportation. (b) $|X^{1/4} \rangle$ state teleportation. (c) $\ket{\text{CC}X}$ state teleportation. All three gates are in the third level of the Clifford hierarchy and allow the implementation of a universal set of gates if supplemented by the Clifford group.}
    \label{fig:magic-state}
\end{figure}

Magic state preparation and teleportation is the state-of-the-art method for performing fault-tolerant non-Clifford gates. High-fidelity magic states can be consumed to implement the corresponding gates, as shown in Figure~\ref{fig:magic-state}.

\textit{Magic state distillation.}---
Magic state distillation is one of the leading approaches for preparing high-fidelity magic states~\cite{bravyi2005universal}. At a high level, the protocol takes low-fidelity magic states as input and produces fewer magic states as output, but with higher fidelities. However, magic state distillation operates at the logical level, and thus comes with a really high cost both in space and time overhead~\cite{litinski2019magic}. One of the most studied protocols for preparing $\ket{T}$ magic states relies on the $[[15, 1, 3]]$ quantum Reed-Muller code which possesses a transversal $T$ gate~\cite{bravyi2005universal}. Magic state distillation then consists in concatenating the Reed-Muller code as the outer code, and the surface codes as the inner codes, thus requiring 15 patches of surface code, explaining the large space overhead.

Magic state distillation can also be used for biased-noise qubits. The overhead reduction for quantum memory is beneficial, as the logical qubits used in the protocol can be implemented with codes tailored to noise bias, as seen in Subsection~\ref{sec:quantum_codes}. Nevertheless, the relative cost of magic state distillation compared to a quantum memory does not change, and the process still demands many logical qubits and considerable time to produce a high-fidelity magic state. Specific features of Kerr cat qubits can however be leveraged to improve the fidelity of the magic states used as input to the magic state distillation protocol. When implementing the gate $ZZ(\theta) = e^{-i\frac{\theta}{2} ZZ}$, a correlated error is quadratically suppressed compared to single-qubit errors $p_{ZZ} \approx p_Z^2$. This property allows for the preparation of magic states with an error rate in $\mathcal O (p_Z^2)$ as demonstrated in Ref.~\cite{singh2022high}. This permits lowering the error rate in the magic state injection protocol. However, for dissipative cat qubits, correlated errors in diagonal gates are limited by non-adiabatic errors, preventing the use of this method.

\begin{figure*}
    \centering
    \includegraphics[width=\textwidth]{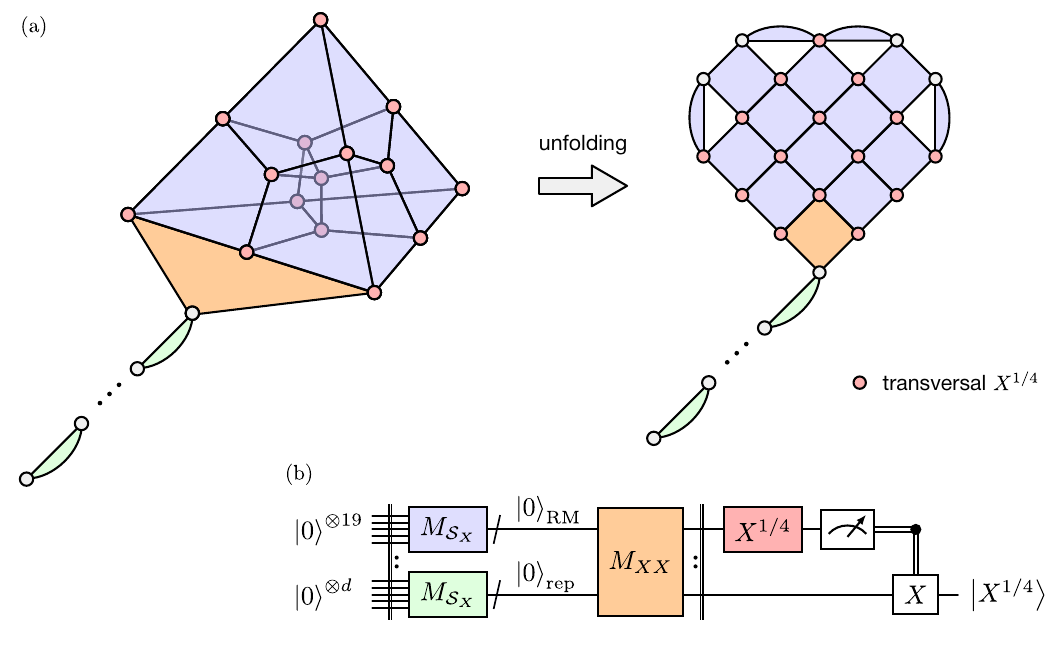}
    \caption{(Figure taken from our earlier publication~\cite{ruiz2025unfolded}) Unfolding of the $X$-type stabilizer group of the quantum Hadamard Reed-Muller code on a 2D layout. (a) Each qubit of the quantum Hadamard Reed-Muller code (left) corresponds to a red qubit in the \textit{unfolded} code (right). To enable nearest-neighbor connectivity, four additional data qubits (in grey) are introduced in the unfolded code, along with weight-2 stabilizers. The repetition code that will host the output magic state (green) is merged with the unfolded code via a joint $X_\text{RM} X_\text{rep}$ measurement (orange). (b) Logical circuit corresponding to the unfolded distillation protocol. It begins by initializing all data qubits in the state $\ket{0}$, followed by the measurement of the $X$-type stabilizers of the unfolded code (blue), the $X_\text{RM} X_\text{rep}$ joint logical operator (orange) and the repetition code stabilizers (green) for several rounds. A transversal $X^{1/4}$ gate is then applied to the red qubits of the unfolded code, followed by the measurement of all data qubits of the unfolded code in the $Z$ basis. The four $Z$ stabilizers of the Hadamard Reed-Muller code are then reconstructed from these outcomes. If all stabilizers yield $+1$, the distillation is considered successful, and the repetition code is projected into the state $|X^{1/4}\rangle$.}
    \label{fig:unfolded_distillation}
\end{figure*}

\textit{Clifford measurement.}---Clifford measurement is an alternative to magic state distillation and has been shown to outperform it for standard qubits down to a logical error rate of $10^{-9}$ when the physical error rate is equal to or lower than $10^{-3}$~\cite{gidney2024magic}. The scheme operates by measuring a logical Clifford operator that is transversal on the given code and stabilizes the target magic state~\cite{shor1996fault,gottesman1999demonstrating, zhou2000methodology}. The recent magic state cultivation construction proposes to measure $H_{XY} = (X+Y)/\sqrt{2}$ on the 2D color code to prepare a $\ket{T}$ state~\cite{zhou2000methodology,gidney2024magic}. To ensure fault tolerance, the Clifford measurement is performed with ancilla qubits prepared in a verified $\ket{GHZ}$ state (or with an in-place folding method with flag qubits for cultivation), preventing error from propagating from the ancilla to the code. However, the $\ket{GHZ}$ state measurement itself is noisy and must be repeated and interleaved with postselected rounds of quantum error correction on the code. 

For biased-noise qubits, it was also proposed to measure $X_1 (CX)_{23}$ on three repetition codes to prepare a $\ket{\text{CC}X} = \tfrac{1}{2}(\ket{000} + \ket{010} + \ket{100} + \ket{111})$ state~\cite{zhou2000methodology,chamberland2022building}. It can be performed  locally between three repetition codes using transversal Toffoli gates~\cite{chamberland2022building,gouzien2023performance}. Interestingly, the parity measurement verification step of the $\ket{GHZ}$ state used to perform the Clifford measurement, unlike in the case of unbiased-noise qubits, can be skipped for biased-noise qubits. Indeed, as bit-flip errors are suppressed, the probability of a high-weight error on the $\ket{GHZ}$ state is suppressed as well. 

Although this method has a very low qubit overhead (one surface code for the $H_{XY}$ measurement and three repetition codes for the $X_1 (CX)_{23}$ measurement),  the postselection process may incur a substantial time overhead. This results in a scheme that imposes stringent requirements on the physical error rate. For instance, for the $H_{XY}$ measurement, increasing the physical error rate from $p = 0.1\%$ to $p = 0.2\%$ results in a $77 \times$ increase in preparation time~\cite{gidney2024magic}. Similarly the $\ket{\text{CC}X}$ scheme is highly sensitive to the physical error rate. Consequently, for dissipative cat qubits, the Clifford measurement scheme imposes a very demanding requirement on $\kappa_1/\kappa_2$, with $\kappa_1$ representing the single-photon loss rate and $\kappa_2$ the two-photon dissipation rate used to stabilize cat qubits (see Figure 15 in Ref.~\cite{chamberland2022building}).

\textit{Unfolded distillation.}---While the previous  methods can be applied to both standard and biased-noise qubits, we now present a new method introduced in Ref.~\cite{ruiz2025unfolded} that specifically leverages the noise bias for preparing magic states. In contrast to the previous schemes, the method prepares a magic state $\ket{X^{1/4}} = \cos(\pi/8)\ket{0} - i \sin(\pi/8) \ket{1}$. This magic state can be used as a resource to apply an $X^{1/4}$ gate, equivalent to the $T$ gate up to Hadamard gates. Like magic state distillation, unfolded distillation relies on a code with a transversal non-Clifford gate, and in this case, we consider the Hadamard Reed-Muller code, where $X$ and $Z$-type stabilizers are switched compared to the standard Reed-Muller code, and which possesses a transversal $X^{1/4}$ gate. Because the Reed–Muller code is a 3D code, it cannot be implemented at the physical level on a 2D qubit architecture. To overcome this limitation, magic state distillation typically employs it at the logical level through code concatenation. However, the presence of noise bias enables a different approach: the Hadamard Reed–Muller code can be used directly at the physical level rather than at the logical level. The key insight is that, due to the noise bias, it is sufficient to repeatedly measure only the $X$-type stabilizers of the code. In this setting, a generating set of the $X$-stabilizer group of the Hadamard Reed–Muller code can be measured while preserving a 2D local architecture, as illustrated in Figure~\ref{fig:unfolded_distillation}(a).

Figure~\ref{fig:unfolded_distillation}(b) illustrates how the unfolded code can be leveraged to prepare a magic state. We first prepare the unfolded code and a repetition code, which will host the output magic state, in the state $\ket{0}_L$. The two codes are then merged through a logical $XX$ measurement, projecting them into a Bell pair. We subsequently apply a transversal $X^{1/4}$ gate to the unfolded code, followed by a $Z$ measurement of the unfolded code. This procedure teleports the $\ket{X^{1/4}}$ state onto the repetition code, up to Pauli corrections. Importantly, the final $Z$ measurement of the unfolded code allows us to reconstruct the $Z$-stabilizer values of the Hadamard Reed-Muller code and detect errors occurring during the application of the $X^{1/4}$ gates. As is usual with distillation protocols, if any error is detected through these $Z$ measurements, the preparation is rejected. We emphasize that the $X^{1/4}$ gate is not bias-preserving and can therefore introduce bit-flip errors. However, these errors are detected by the $Z$ stabilizers extracted from the subsequent measurement. 

Ref.~\cite{ruiz2025unfolded} demonstrated that a magic state with a logical error rate of $6 \times 10^{-8}$ can be prepared using only 53 qubits, requiring an average of 12.2 rounds of stabilizer measurements, at a physical error rate of $p_Z = 10^{-3}$. This relies on a large noise bias of $5 \times 10^7$, as the output repetition code provides no protection against bit-flip errors. However, the scheme can be made more robust to bit-flip errors by merging the unfolded code with a thin surface code. The resulting output patch is then protected against bit-flip errors, while, remarkably, the final measurement of the $Z$ stabilizers of the Hadamard Reed-Muller code can also detect bit-flip errors occurring within the unfolded code. Consequently, the scheme remains efficient even at significantly lower noise bias. Ref.~\cite{ruiz2025unfolded} showed that a magic state with a logical error rate of $7 \times 10^{-7}$ can be prepared with a noise bias of $\eta = 80$ and a physical error rate of $p_Z = 10^{-3}$ with only 175 qubits and an average of 9.6 rounds. At the higher error rate of $p_Z = 5\times 10^{-3}$, it is still possible to prepare a magic state with a logical error rate of $2 \times 10^{-5}$ using only 16.6 rounds of stabilizer measurements on average (this value is for the repetition-code scheme and might change slightly for the surface-code scheme). Although this error rate is not sufficiently low for direct use in quantum algorithms, the resulting magic state can serve as an input to subsequent standard magic state distillation protocols, such as an $8T \rightarrow \mathrm{CCZ}$ distillation protocol, enabling the implementation of Toffoli gates with a logical error rate of approximately $10^{-8}$~\cite{gidney2019efficient}. More interestingly, it is also possible to extend the unfolding idea to codes with higher distances that admit transversal non-Clifford gates. This is shown for instance in~\cite{londe-2026} with a 2D unfolding of a Reed-Muller code of distance 4. An extension to even higher distances could remove the requirement for concatenation  with a second round of standard distillation.

A similar method can be leveraged to prepare $\ket{+i}$ states at a low cost, required to complete the implementation of the Clifford group. Instead of merging a repetition code or a thin surface code to the unfolded Hadamard Reed-Muller code, we could simply merge them to a 2D color code, that possesses a transversal $S$ gate~\cite{ruiz2025unfolded}. Then, the exact same protocol as the one presented in Figure~\ref{fig:unfolded_distillation} can be used. We note that because the 2D color code has fewer qubits than its 3D counterpart, the protocol has a smaller qubit overhead. Second, because the code is already two-dimensional, no unfolding is required, and the method can be straightforwardly extended to larger-distance 2D color codes.

\section{\label{sec:ZZZ meas}Measurement-based architecture}

We saw in Section~\ref{sec:CZ} that the absence of bit-flip errors does not have a significant impact on the hardware overhead of error correction when the set of bias-preserving operations is limited to $\{\text{C}Z,\cP_{\ket{+}},\cM_X\}$. Indeed, while the correction of phase-flip errors require the measurement of $X$-type stabilizers, the gadgets that ensure such measurements with the above elementary gate set are rather complex. This complexity strongly limits the performance of the error correction. 

As seen in Section~\ref{sec:CNOT}, the situation changes drastically with the addition of the bias-preserving C\(X\). At the quantum memory level, this opens up room for significant overhead savings with two scenarios: 1- With moderate noise bias of order $10^3-10^4$ or less, the optimal strategy could be to use a rectangular XZZX code or the concatenated codes, based on the desired logical error rate ; 2- With larger noise biases, the best option is always the concatenated approach where we correct for phase-flips and take care of bit-flips through concatenation with a high-rate bit-flip code. In order to extend this overhead reduction to fault-tolerant operations, we further need to implement high-fidelity preparation $\cP_{\ket{0}}$ and measurement $\cM_Z$, compatible with both naturally and engineered biased-noise qubits. 

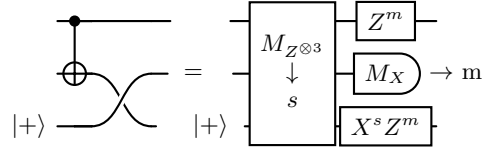
\begin{figure}[h!]
    \centering
    \vspace{1em}
    \begin{quantikz}[row sep=0.5em, column sep=0.2em]
	                    \lstick{\hspace{-3em}}&\ctrl{1}&\permute{1,3,2}& \midstick[1,brackets=none]{$\;\;\;\;$}         & \gate[3, disable auto height][1][1]{\begin{matrix}
	                   M_{Z^{\otimes 3}}\\ 	\downarrow\\s
	               \end{matrix}}&\gate{Z^m}& \\
	                      \lstick{\hspace{-3em}}&\targ{} &               & \midstick[1,brackets=none]{$=\;\;$}&&                   \meterD{M_X}\rstick{$\rightarrow$ m} \\
\lstick{\hspace{-3em}$\qquad\ket{+}$}&        &               & \midstick[1,brackets=none]{$\;\;\;\ket{+}$}&                                      &\gate{X^sZ^m}&
    \end{quantikz}
    \caption{Circuit to implement a C\(X\) using a QND \(Z^{\otimes 3}\) measurement. Depending on the \(Z^{\otimes 3}\) measurement outcome there is an \(X\) correction to be applied. This correction can be kept in a Pauli frame if one does not have a physical implementation of a bias preserving \(X\) gate. Moreover, the assignment error of the \(Z^{\otimes 3}\) measurement has to be as small as the bit-flip error probability for the protocol to preserve the bias.}
    \label{fig:CX}
\end{figure}

As discussed in Section~\ref{sec:bias}, while such bias-preserving C\(X\) is usually not implementable with naturally biased-noise qubits, its implementation with engineered ones also represents some difficulties. In this Section, we discuss an alternative solution presented in~\cite{vuillot2026}. In this approach, the main ingredient of the architecture shifts from bias-preserving C\(X\) to high-fidelity QND measurement of weight-3 or weight-4 Pauli \(Z\) operators. Indeed, the main idea lies in the equivalence between the logical circuits represented in Fig.~\ref{fig:CX}. Similarly to the case of equivalence in Fig.~\ref{fig:MXX}, here we do not need to actually implement the feedforward Pauli $X$ and $Z$ gates, and we can simply keep track of them in the Pauli frame. However, to ensure the bias-preserving nature of the operation, we need to ensure an ultra-high-fidelity QND readout of the $Z^{\otimes 3}$ operator as any assignment error leads to an $X$ error (a bit-flip) in the Pauli frame. As briefly discussed in Section~\ref{sec:bias} and in~\cite{vuillot2026}, there are ways to implement such high-fidelity readout in both naturally and engineered biased-noise qubits. All the constructions of the previous section with bias-preserving C\(X\) can be transposed into this new case. In this section, we briefly review the implementation of error correction with this central ingredient. We also provide a few comments on the transposition of the constructions for fault-tolerant operations 
introduced in the previous Section.

As discussed in the previous Section, the correction of phase-flips requires the measurement of \(X\) stabilizers in case of significant noise bias or mixed $XZZX$ stabilizers in case of moderate noise bias. Let us start by discussing the measurement of weight-2 $X$ stabilizers for a phase-flip repetition code in the case of significant noise bias. Noting that the $XX$ parity can be mapped to a measurement ancilla qubit using two C\(X\) operations, the transposition of this circuit using the equivalence of Fig.~\ref{fig:CX} would require 2 additional qubits, the implementation of 2 weight-3 QND \(Z\) readout followed by 3 destructive \(X\) readouts. This circuit can be recompiled using weight-4 QND \(Z\) readout and would reduce to the circuit presented in Fig.~\ref{fig:MXX}. In Subsection~\ref{ssec:knill}, we discussed the phase-flip error correction using this equivalence but we considered the case where such a high-fidelity QND readout of weight-4  \(Z\) operators is performed using C\(Z\) gates. Noting that the ultra-high-fidelity readout required for this scheme is achieved by repeating several rounds such a measurement, the performance of this approach remains quite limited as seen in Fig.~\ref{fig:CZ_overhead} (blue curve). 

In~\cite{vuillot2026}, we argued that the real strength of the multi-Z measurement primitive for hardware savings relies on the fact that the measurement device does not need to be based on the same physical qubits as the ones used for computation. Indeed, in this same reference, we discussed a few  efficient approaches to realize such readout using e.g. electron spins  coupled to nuclear spins, or by a continuous parametric readout for cat qubits. In this subsection, we review the performances achieved with this primitive and compare them with the performances of Section~\ref{sec:CNOT}.

\begin{figure}[t]
    \centering
        \vspace{2em}
    \raisebox{7em}{\resizebox{\linewidth}{!}{\begin{quantikz}[row sep=0.5em]
			\lstick{$\ket{+}$} & \gate[4, disable auto height][1][1]{\begin{matrix}
					M_{Z^4}\\ 	\downarrow\\s_1
			\end{matrix}}&&\gate{Z^{m_1}}&&&\gate{X^{s_1+s_2+s_3+s_4}}& \\
			\lstick{$D_0$} &&\meterD{M_X}&\setwiretype{n} \midstick{\hspace{-2.3em}$=m_1\quad\ket{+}$}&\setwiretype{q}&&&\\
			\lstick{$\ket{+}$} &&&\gate{Z^{m_2}}&\gate[4, disable auto height][1][1]{\begin{matrix}
					M_{Z^4}\\ 	\downarrow\\s_2
			\end{matrix}}\setwiretype{q}& \meterD{M_X}&\setwiretype{n}\midstick{\hspace{-5.1em}$= m_5\quad \ket{+}$}&\setwiretype{q}\\
			\lstick{$D_1$} &&\meterD{M_X} &\setwiretype{n}\midstick{\hspace{-2.3em}$=m_2\quad\ket{+}$}&\setwiretype{q}&&\gate{Z^{m_5}}&\\
			\lstick{$\ket{+}$} & \gate[4, disable auto height][1][1]{\begin{matrix}
					M_{Z^4}\\ 	\downarrow\\s_3
			\end{matrix}}&&\gate{Z^{m_3}}&& \meterD{M_X}&\setwiretype{n}\midstick{\hspace{-5.1em}$= m_6\quad \ket{+}$}&\setwiretype{q}\\
			\lstick{$D_2$} &&\meterD{M_X} &\setwiretype{n}\midstick{\hspace{-2.3em}$=m_3\quad\ket{+}$}&\setwiretype{q}&&\gate{Z^{m_6}}&\\
			\lstick{$\ket{+}$} &&&\gate{Z^{m_4}}&\gate[4, disable auto height][1][1]{\begin{matrix}
					M_{Z^4}\\ 	\downarrow\\s_4
			\end{matrix}}\setwiretype{q}& \meterD{M_X}&\setwiretype{n}\midstick{\hspace{-5.1em}$= m_7\quad \ket{+}$}&\setwiretype{q}\\
			\lstick{$D_3$} &&\meterD{M_X} &\setwiretype{n}\midstick{\hspace{-2.3em}$=m_4\quad\ket{+}$}&\setwiretype{q} &&\gate{Z^{m_7}}&\\
			\lstick{$\ket{+}$} &&&&&&\gate{Z^{m_8}}& \\
			\lstick{$D_4$} &&&&& \meterD{M_X}&\setwiretype{n}\midstick{\hspace{-5.1em}$= m_8\quad \ket{+}$}&\setwiretype{q}\\
		\end{quantikz}}}
    \caption{%
        Logical circuit for the phase-flip repetition code syndrome extraction (alternating schedule, $d=5$). This represents a single cycle, extracting the full
        syndrome $\sigma = (m_1\oplus m_2,\, m_3\oplus m_4,\, m_5\oplus m_6,\, m_7\oplus m_8)$.
    }
    \label{fig:alternated}
\end{figure}
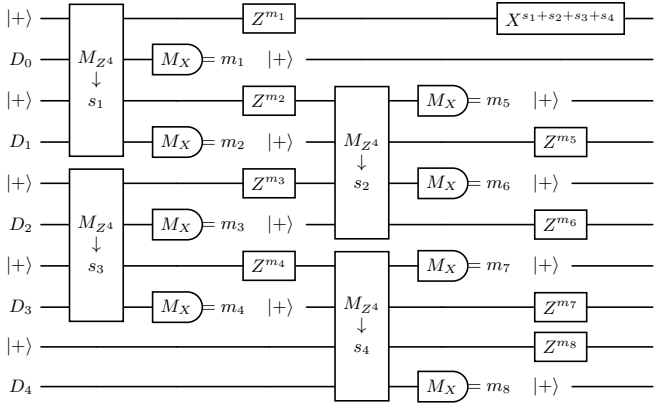

Although other gate schedules are possible and discussed in~\cite{vuillot2026}, here we focus on the alternate schedule as shown in Fig.~\ref{fig:alternated}. This scheduling has the merit of not increasing the number of ancilla qubits with respect to the C\(X\)-based phase-flip repetition code presented in Section~\ref{ssec:phase-flip}. However, the circuit depth of this extraction cycle  increases from 4 time-steps in the C\(X\)-based scheme to 6. Once again, we also note that the Pauli $X$ and $Z$ gates presented in the figure do not need to be applied in practice and can be taken into account in the Pauli frame. Finally, note that at the end of each syndrome extraction cycle, the data qubits are teleported and after two cycles they are back to their initial place.

\begin{figure}
    \centering
    \includegraphics[width=\columnwidth]{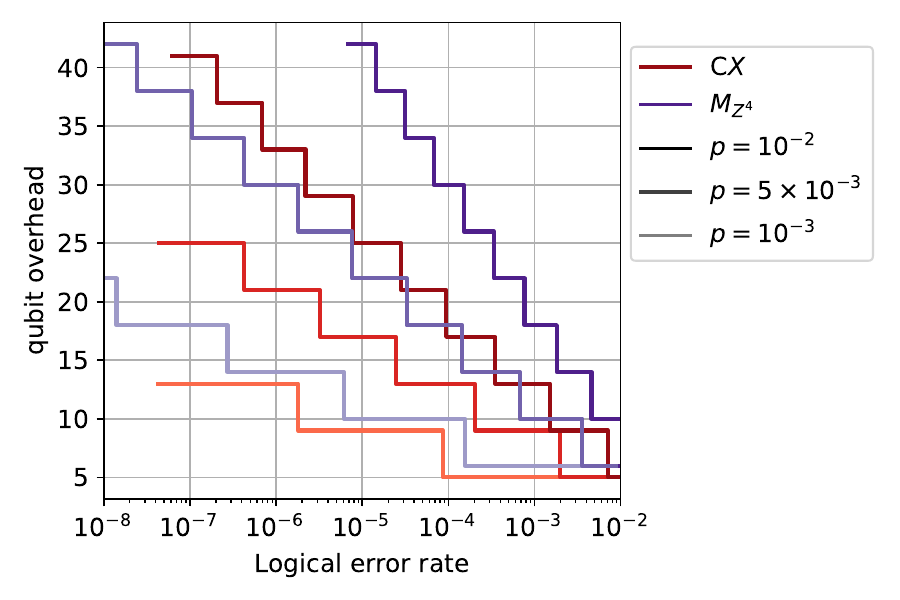}
    \caption{Qubit overhead comparison for the repetition code at infinite noise bias.
    For each target logical error rate per cycle (x-axis), the minimum number of physical qubits is shown for the error correction circuit of Fig.~\ref{fig:alternated} with $M_{Z^4}$ primitive (purple curves). 
    For comparison, we also provide the overhead associated to the error correction circuit of Fig.~\ref{fig:repetition code} with the bias-preserving CNOT primitive (red curves). 
    }
    \label{fig:rep_code_MZ}
\end{figure}

In Fig.~\ref{fig:rep_code_MZ}, we present the result of numerical simulations for this error correction strategy (purple curves). 
In these simulations, we assume the circuit-level error model given in Table~\ref{tab:noise-model_MZZ}. While single-qubit operations and idle, as well as preparation and measurement in the $X$ basis, are submitted to the same error model as in previous sections, the new primitive associated with the QND measurement $\cM_{Z^{\otimes 4}}$, suffers from a phase-flip type error of probability $2p_Z$, shared equivalently between the four involved qubits. This error rate is twice that of the bias-preserving C\(X\) gate. Indeed, noting that this measurement involves twice more qubits than  C\(X\) and actually replaces two bias-preserving C\(X\) operations in the error correction circuit, we have  made this choice of the error model to ensure a fare comparison between the architectures. Finally, we note that the assignment error probability of this QND measurement matches the bit-flip probability. Finally, we also have assumed that this QND measurement is accompanied by uncorrelated single-qubit $X$- and $Y$-type errors of probability $2p_X$ in total. Note that this choice does not have any impact on the performance of the phase-flip repetition code that does not correct any bit-flips.

\begin{table}[htbp]
\renewcommand{\arraystretch}{1.6} 
\begin{ruledtabular}
\begin{tabular}{lcc}
Operation & Error type & Probability \\
\colrule
\multirow{2}{*}{Single-qubit \& Idling} & $X, Y$ & $\frac{p_X}{2}$ \\
                       & $Z$    & $p_Z$ \\
\colrule
Preparation $\cP_{\ket{+}}$ & $Z$    & $p_Z$ \\
\colrule
Meas $\cM_{X}$ & assignment error    & $p_Z$ \\
\colrule
Two-qubit C\(Z\)  & ZI,IZ    & $\frac{p_Z}{2}$ \\
                       &$P_1 P_2$, $P_1 \text{ xor } P_2 \in \{X,Y\}$ & $\frac{p_X}{8}$ \\
\colrule
QND $\cM_{Z^{\otimes 4}}$  & Single-qubit $Z$  & $\frac{p_Z}{2}$ \\
                      & Single-qubit $X$ and $Y$   & $\frac{p_X}{4}$ \\
                       & assignment error    & $p_X$ \\
\end{tabular}
\end{ruledtabular}
\caption{\label{tab:noise-model_MZZ} The error model for the error correction circuit of Figs.~\ref{fig:alternated} and~\ref{fig:XZZX-schedule-MZ4}.}
\end{table}

For comparison, we also provide in the same Figure the overhead with bias-preserving C\(X\) gates instead of weight-4 Pauli $Z$ measurements (red curves). As it can be observed the error correction with bias-preserving C\(X\)'s perform slightly better. This is mainly explained with the slightly better error correction threshold due to shallower syndrome extraction cycles. 

\begin{figure}
    \centering
    \includegraphics[width=\columnwidth]{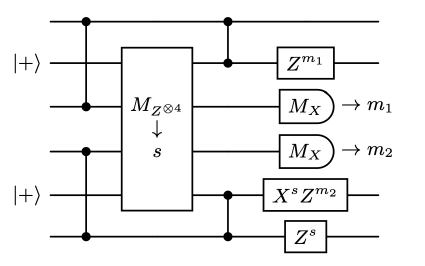}
    \caption{Circuit to implement the mixed parity check $ZXXZ$ using a single
    QND $\cM_{Z^{\otimes 4}}$ measurement and 4 bias-preserving C\(Z\) gates. The measurement outcome is given by $m=m_1\oplus m_2$. }
    \label{fig:XZZX-MZ4}
\end{figure}

Now assuming a  moderate noise bias, we discuss the implementation of an asymmetric $XZZX$ code using the primitive of QND $\cM_{Z^{\otimes4}}$. Indeed, the circuit of Fig.~\ref{fig:XZZX-MZ4} realizes such a mixed parity readout using the $\cM_{Z^{\otimes 4}}$ primitive. This gadget uses two ancilla qubits prepared in $\ket{+}$ and also requires 4 bias-preserving C\(Z\) gates. Once again the Pauli corrections could be tracked by updating the Pauli frame. 

\begin{figure}[h!]
\centering
\includegraphics[width=\columnwidth]{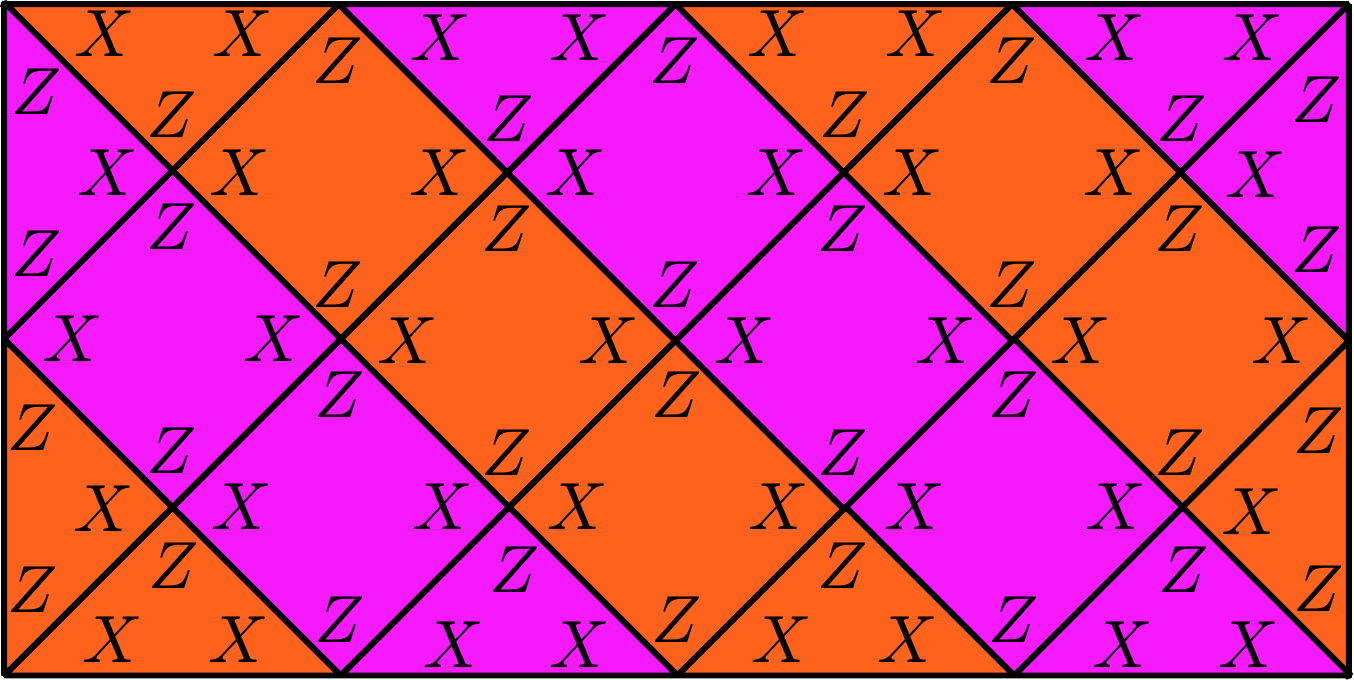}\\
\includegraphics[width=\columnwidth]{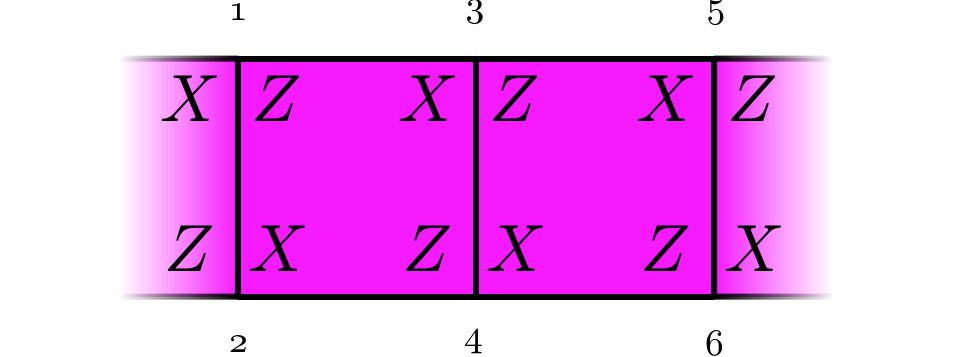}\\[1em]
    \raisebox{7em}{\resizebox{.9\linewidth}{!}{\begin{quantikz}[row sep=0.5em, column sep=0.5em]
        \lstick{$\vdots$}&\setwiretype{n}&\setwiretype{q}& \phase{}&&\cdots\setwiretype{n}&&\ghost{X}\setwiretype{q}& \\
        \lstick[2]{1} &\setwiretype{n}&\ctrl{2}\setwiretype{q}&&&\cdots\setwiretype{n}&\ghost{X}& \setwiretype{q}&\\
        &\setwiretype{n}\midstick{$\ket{+}$} &\setwiretype{q}&\ctrl{2}&&\cdots\setwiretype{n}&\ghost{X}&\setwiretype{q}&\\
        \lstick[2]{2} &\setwiretype{n}&\ctrl{-2}\setwiretype{q}&&& \gate[4, disable auto height][1][1]{\begin{matrix}
                M_{Z^4}\\ 	\downarrow\\s_1
        \end{matrix}}&\meterD{M_X}\rstick{\hspace{-.45em}$= m_1$} \\
        &\setwiretype{n}\midstick{$\ket{+}$} &\setwiretype{q}&\ctrl{-2}&&&&\gate{Z^{m_1}}&\\
        \lstick[2]{3} &\setwiretype{n}&\ctrl{2}\setwiretype{q}&&&& \meterD{M_X}\rstick{\hspace{-.45em}$= m_2$}\\
        &\setwiretype{n}\midstick{$\ket{+}$} &\setwiretype{q}&\ctrl{2}&&&&\gate{X^{s_1}Z^{m_2}}&\\
        \lstick[2]{4} &\setwiretype{n}&\ctrl{-2}\setwiretype{q}&&&\gate[4, disable auto height][1][1]{\begin{matrix}
                M_{Z^4}\\ 	\downarrow\\s_2
        \end{matrix}}& \meterD{M_X}\rstick{\hspace{-.45em}$= m_3$}\\
        &\setwiretype{n}\midstick{$\ket{+}$} &\setwiretype{q}&\ctrl{-2}&&&& \gate{Z^{m_3+s_1}}&\\
        \lstick[2]{5} &\setwiretype{n}&\ctrl{2}\setwiretype{q}&&&&\meterD{M_X}\rstick{\hspace{-.45em}$= m_4$}\\
        &\setwiretype{n}\midstick{$\ket{+}$} &\setwiretype{q}&\ctrl{2}&& &&\gate{X^{s_2}Z^{m_4}}& \\
        \lstick[2]{6} &\setwiretype{n}&\ctrl{-2}\setwiretype{q}&&&\cdots\setwiretype{n}&\ghost{X}& \setwiretype{q}&\\
        &\setwiretype{n}\midstick{$\ket{+}$} &\setwiretype{q}&\ctrl{-2}&&\cdots\setwiretype{n}&&\gate{Z^{s_2}} \setwiretype{q}&\\
        \lstick{$\vdots$}&\setwiretype{n}&\phase{}\setwiretype{q}&&&\cdots\setwiretype{n}&&\ghost{X}\setwiretype{q}& \\
    \end{quantikz}}}
\caption{Gate schedule for the error syndrome extractions of the $XZZX$ code with the primitive $\cM_{Z^{\otimes 4}}$. The stabilizers are partitioned in two parts (orange and purple) and are measured on alternate rounds. Each vertex of this lattice hosts two physical qubits. The circuit corresponds to the realization of one such round.  }
\label{fig:XZZX-schedule-MZ4}
\end{figure}

One possible scheduling of the XZZX error correction is introduced in~\cite{vuillot2026}. We recall it in Fig.~\ref{fig:XZZX-schedule-MZ4}  for the sake of completeness. The stabilizers are measured in two rounds following the logical circuit provided in this Figure. Also note that, this scheduling requires an ancilla qubit per data qubit, so no more spatial overhead than the realization of the $XZZX$ code with bias-preserving C\(X\) operations. 

\begin{figure}
\centering
\includegraphics[width=\columnwidth]{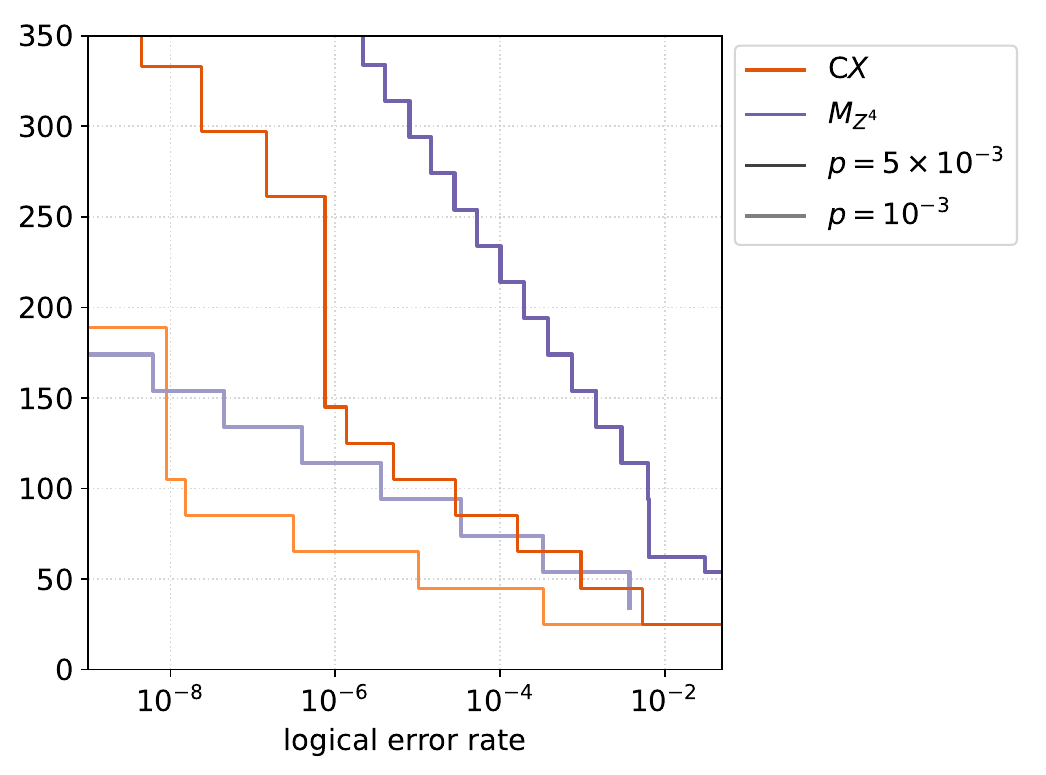}
\caption{Qubit overhead comparison for the $XZZX$ code implemented using QND $\cM_{Z^{\otimes 4}}$ (purple curves) as in Figure~\ref{fig:XZZX-schedule-MZ4} or bias preserving C$X$ gates (orange curves) as in Figure~\ref{fig:XZZX code}.
We choose a moderate bias of $\eta=10^3$ and two different $p_Z\in\{10^{-3}, 5\cdot 10^{-3}\}$ with the lighter curves being the smaller $p_Z$.
\label{fig:sim_xzzx_mz4}
}
\end{figure}

The performance of this realization is illustrated in Fig.~\ref{fig:sim_xzzx_mz4}. In these simulations, we consider again the error model detailed in Table~\ref{tab:noise-model_MZZ}. Note that while no correlations in bit-flip type errors are considered for the $\cM_{Z^{\otimes 4}}$, we include such correlated errors in the two-qubit C\(Z\) gates. 
For comparison, we also provide the overheads associated to the same realization of the $XZZX$ code with the bias-preserving C\(X\) primitive instead of $\cM_{Z^{\otimes 4}}$.
We can see that the overheads are comparable and even close when $p_Z$ is smaller.

We finish this section with a brief discussion of fault-tolerant operations with the $\cM_{Z^{\otimes 4}}$ primitive. As discussed in Section~\ref{ssec:cx_ft}, with the elementary set of bias-preserving operations $\{
\cP_{\ket{+}},\cP_{\ket{0}},\cM_X,\cM_Z,\text{C}Z,\text{C}X\}$
and the bias-non-preserving operations $X^{\pm1/4}$, we are able to construct in hardware-efficient manner all the operations of a universal set of fault-tolerant gates. Noting the equivalence of Fig.~\ref{fig:CX}, we can simply replace in all these constructions the bias-preserving C\(X\) with high-fidelity QND measurements $\cM_{Z^{\otimes 3}}$. Also note that a high-fidelity QND measurement $\cM_Z$ also ensures a bias-preserving preparation $\cP_{\ket{0}}$. So in this new architecture, we can replace the elementary set of operations by
$$
\{\cP_{\ket{+}},\cM_{X},\text{QND-}\cM_Z,\text{QND-}\cM_{Z^{\otimes3}},\text{C}Z\}_{\text{bp}}\cup\{X^{\pm1/4}\}_{\text{bnp}}
$$
where bp (resp. bnp) stands for bias-preserving (resp. bias-non-preserving). 

\section{Conclusion}\label{sec:conclusion}

In this review, we provided a thorough analysis of state-of-art approaches to build a hardware-efficient fault-tolerant quantum processor based on biased-noise qubits. We discussed the two cases of natural noise bias such as electron or nuclear spins and the engineered noise bias such as the stabilized cat qubits. We argued that these cases  differ in respect with the set of bias-preserving physical operations, i.e. operations that could be implemented without converting the frequent phase-flips to rare bit-flips. We then discussed three configurations based on the set of elementary physical operations  that could be implemented in a bias-preserving manner. 

In the case where this set is reduced to preparation and measurement in the $X$ basis and the gates that are diagonal in the computational basis such as the two-qubit C\(Z\) gate, we showed that the expected overhead reduction for quantum error correction is very slim at best. More precisely, at physical noise levels that are achievable with current experiments, the absence of bit-flip errors does not significantly reduce the overhead of error correction compared to standard QEC approaches with depolarizing noise models of the same rate. 

The real interest of the biased noise is unleashed with the addition of a bias-preserving C\(X\) gate. We analyzed various QEC strategies in presence of such a gate and we demonstrated that the best performances are achieved when we handle the frequent  phase-flip errors through an appropriate phase-flip code and then concatenate with a high-rate bit-flip code to correct for such rare errors. We also discussed the strategies to implement fault-tolerant gates and demonstrated that the preparation of high-fidelity magic states can be drastically simplified with such biased noise. In this regard, we note two directions for the possible improvements. First, the fault-tolerant gate strategies presented in this review apply to phase-flip codes as well as the $XZZX$ or thin surface codes. An extension to the case of concatenated elevator codes that benefit from the best QEC performances in regard with the qubit  overhead is therefore still missing. Second, the unfolded distillation approach that represent the best overhead performances in regard with high-fidelity magic state preparation ensures a fist level of distillation and to reach ultra-high fidelities one might need to concatenate this approach with another level of standard distillation. It should however be possible to extend this unfolding approach to higher-distance codes with transversal non-Clifford gates and therefore remove this requirement for the second level of distillation.  

Noting that such bias-preserving C\(X\) gates are forbidden in most naturally biased-noise platforms and are complex to implement in engineered biased-noise ones, we consider a third configuration where further to the diagonal gates we have access to a high-fidelity quantum non-demolition readout of multi-qubit Pauli $Z$ operator. We argue that this primitive could replace the bias-preserving C\(X\) in all previous constructions and as such provide similar overhead reductions for quantum error correction and fault-tolerance.

\end{document}